\documentclass[%
 reprint,
nofootinbib,
 amsmath,amssymb,
 aps,
floatfix,
]{revtex4-2}

\usepackage{graphicx}% Include figure files
\usepackage{dcolumn}% Align table columns on decimal point
\usepackage{bm}% bold math
\usepackage{xcolor}
\begin{document}

\preprint{APS/123-QED}

\title{Growth-Induced Transitions in Viscoelastic Matter}% Force line breaks with \\

\author{Valentin Slepukhin}
 \altaffiliation{Leipzig University.}%Lines break automatically or can be forced with \\
 \email{valentin.slepukhin@uni-leipzig.de}
\author{Oskar Hallatschek}%
 \email{ohallats@berkeley.edu}
\altaffiliation{
UC Berkeley}

\date{\today}% It is always \today, today,
             %  but any date may be explicitly specified

\begin{abstract}
Growth is a fundamental process in living systems. Although the stress–deformation response of growing materials is often described as either purely elastic or purely viscous, many biological tissues—from biofilms to tumors—exhibit both elastic and viscous behavior. Here, we show that this viscoelastic response can crucially control the mechanics of proliferating matter when the growth rate becomes comparable to the rate of stress relaxation. Focusing first on the prototypical case of a growing elastic beam, we find that the dynamics are governed by a single dimensionless parameter, $g \tau$, where $g$ is the growth rate and $\tau$ is the viscoelastic relaxation time. While the limits $g \tau \to 0 $ and $g \tau \to \infty$  recover purely viscous and purely elastic behavior, respectively, the intermediate regime is not merely a smooth crossover between them. Instead, qualitatively new dynamics emerge at $ g \tau \sim 1$ , including rapid transitions between metastable states that occur in neither limiting regime. We then develop a general, growth-compatible theoretical framework in which unconstrained growth is intrinsically stress-free, extending the analysis to other prototypical geometries and enabling simulations of more realistic growing biological materials. Within this framework, sharp mechanical transitions arise when stress generated by exponential growth accumulates faster than it can be dissipated by viscoelastic relaxation.
\end{abstract}

%\keywords{Suggested keywords}%Use showkeys class option if keyword
                              %display desired
\maketitle

%\tableofcontents

\section{Introduction}
All living organisms experience growth at certain stages of their development \cite{Hallatschek2023Proliferating}. This includes not only tissue growth during embryogenesis \cite{Lecuit2007} and later developmental stages, but also processes such as wound healing \cite{Martin2015}, tumor expansion \cite{Hanahan2011}, and the growth of microbial communities \cite{kayser2019, Karita2022}. All these phenomena involve proliferating matter.

During growth, proliferating matter is often subjected to stresses arising from deformation. Across different modeling frameworks, the relationship between stress and deformation is treated differently. For instance, organ development has been modeled as growing elastic matter \cite{rodriguez:94}. Biofilms, on the other hand, have been described both as elastic \cite{fei2020} and viscous materials \cite{Giometto2018}, the latter governed by the Stokes equations for viscous fluids. However, even in the absence of growth, it is well known that most biological tissues are neither purely viscous nor purely elastic, but instead exhibit viscoelastic behavior.

How do growth and viscoelasticity interplay? Unconstrained uniform growth with abundant nutrients leads to an exponential increase in the amount of matter, where the growth rate $g$ controls the exponent. If the boundary conditions constrain the growth, this would lead to the exponential buildup of the stress. Viscoelasticity, however, leads to exponential stress decay with a characteristic timescale $\tau$---the viscoelastic relaxation time. The interplay of these two exponential behaviors leads to qualitatively different results than either the purely elastic or purely viscous case. The dynamics of such a material is governed by a single dimensionless parameter $g \tau$. When this parameter is close to zero, the system is in the viscous limit; when it is very large, the system is in the elastic limit.

Naively, one might expect the intermediate regime $g\tau \approx 1$ to represent a smooth crossover between the viscous and elastic limits. This is not the case. We demonstrate that, rather than a gradual transition, the interplay between growth and viscoelasticity can produce a discontinuous jump from viscous-like to elastic-like behavior as $g\tau$ crosses a critical value---more reminiscent of a phase transition: below the critical value, stress decays; above it, stress grows without bound. Moreover, in certain geometries we identify a third regime at intermediate values of $g\tau$ that is qualitatively distinct from both the viscous and elastic limits. In this regime, the interplay of growth and viscoelasticity gives rise to metastable solutions with rapid snaps from one configuration to another.

We organize the rest of the paper as follows. We begin with a system in which the stress remains small, so that the material deformation stays close to the unconstrained-growth scenario---the bending of a growing viscoelastic beam. We show that already in this setting, metastable states arise in the region $g\tau \approx 1$, accompanied by rapid transitions to the stable configuration. To study the more general situation in which stress-induced and growth-induced local deformations\footnote{The total deformation can be big even for the small local deformation, such as for the case of the bending slender beam where each element experiences small deformation, while the whole beam experiences significant bending} can be of the same order of magnitude, we construct a general model for growing viscoelastic matter that exactly recovers the previously studied cases of viscous growth, elastic growth, and viscoelastic non-growing behavior. We demonstrate that, for a simple geometric constraint, the solution exhibits a clear phase transition at a critical value of $g\tau$. Building on this geometry, we explore two more biologically relevant scenarios: growth in a porous medium and wrinkling of a layer of viscoelastic material. In the former, we find that viscoelasticity combined with growth increases the pore permeability, and that the effect arises only from the combination of these two factors. In the latter, we observe that the wrinkling amplitude is maximal in the intermediate regime $g\tau \approx 1$, providing another example in which the intermediate value of $g\tau$ is not simply a smooth interpolation between the viscous and elastic limits.

\section{Dynamics of a viscoelastic growing beam}
\label{sec:unidir}

When a material undergoes uniform growth, no stress should arise unless it is imposed externally---otherwise it would relax away\footnote{The cases when the growing material is embedded in (or intermixed with) the non-growing, such as viscous fluid or polymer matrix, are considered here as stress imposed to the growing material externally by surrounding media. See more in the Discussion section. } .
This is the essential distinction between growth and stretching: the former produces deformation without stress, while the latter generates stress, even if the resulting deformation is the same.

A common pattern in nature is growth that is predominantly one-dimensional: an object elongates while its cross-section remains essentially unchanged\footnote{Note that the full deformation may happen in 3D  due to e.g. bending}. Examples include pollen tubes in plants \cite{Williams2016}, fungal hyphae \cite{Steinberg2017, Brun2023}, filamentous cyanobacteria such as spirulina \cite{Soni2022}, the axons of neurons \cite{Oliveri2022}, and growing microtubules \cite{Gardner2015}. When such growth is unconstrained, the object elongates freely, remaining straight and stress-free.

What happens, however, when the ends of the growing object are pinned? A unidirectionally growing beam whose endpoints are fixed must bend to accommodate the increase in its length. Although the resulting displacement can be large if growth is substantial, the local deformation may nevertheless remain small provided the beam is sufficiently thin compared to its bending curvature radius. For simplicity, from now on we restrict our analysis to planar (two-dimensional) bending to explore how the beam shape changes while it grows. We also neglect the friction to the surrounding media, as well as inertia, assuming them to be significantly smaller than the forces arising due to the constrained growth (the assumption that, as we will see, can not be always satisfied).

\begin{figure*}
    \centering
    \includegraphics[width=0.95\linewidth]{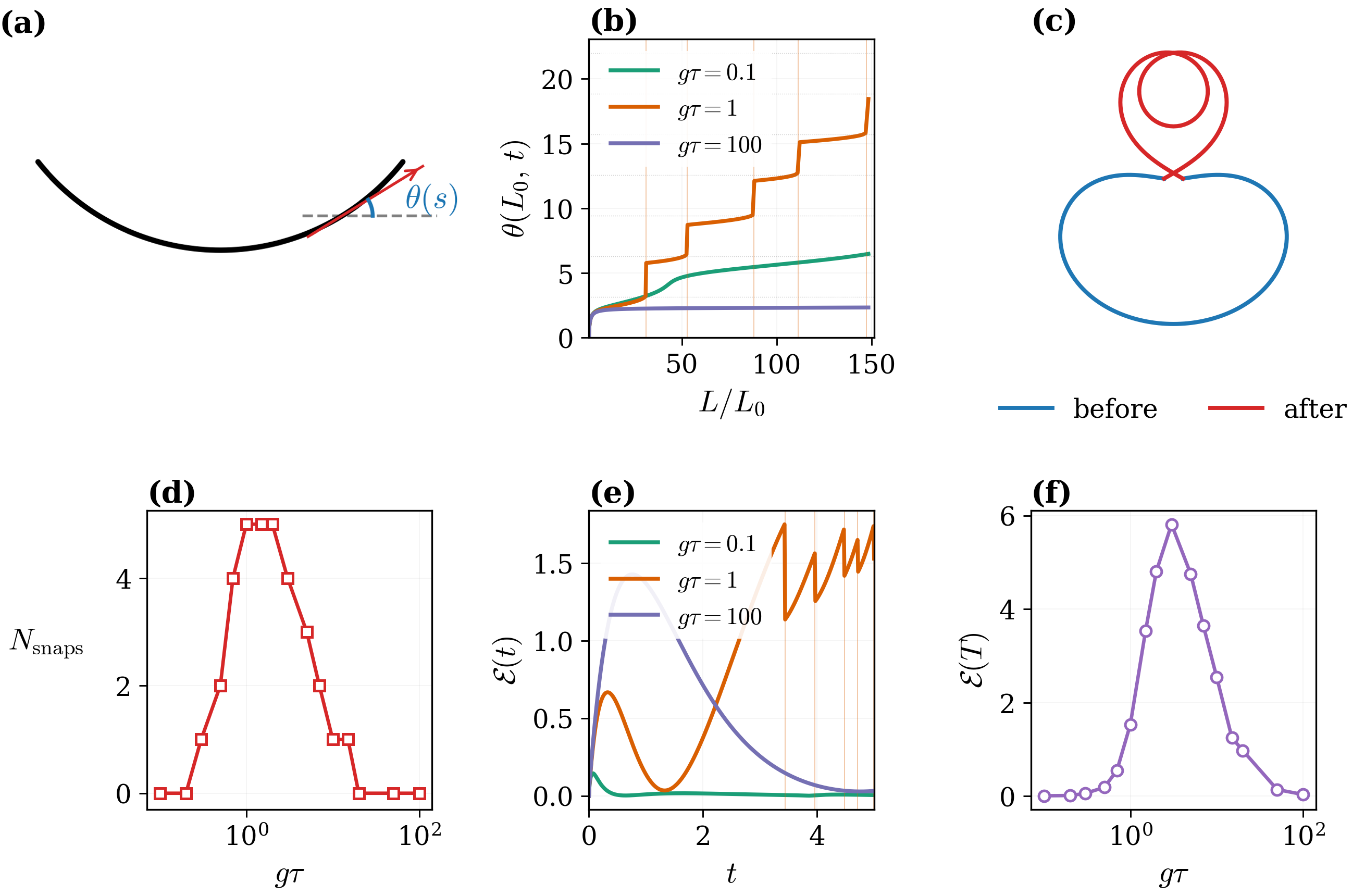}
    \caption{Buckling of a growing viscoelastic beam. (a)~Schematic of the bending deformation. The dashed line indicates the center line; the $x$-axis is aligned with the center line of the straight beam. (b)~Tip angle $\theta(L_0,t)$ as a function of the beam length $L/L_0$. In the elastic case $g \tau \gg 1$ (purple, $g\tau = 100$) the tip angle stops growing at $\theta \approx 140$ degrees. In the viscous regime $g \tau \ll 1$ (green, $g\tau = 0.1$) the angle grows smoothly. In the intermediate case $g \tau \approx 1$ (orange, $g\tau = 1$) the angle demonstrates a ladder pattern with discontinuous jumps by an amount between $\pi/2$ and $\pi$ as the beam snaps --- instantaneously changes its shape. Vertical lines mark the snaps. (c)~Beam profile immediately before (blue) and after (red) a snap, at $g\tau = 1$. (d)~Number of snaps, defined as jumps of the tip angle by $|\Delta \theta| > \pi/2$, accumulated up to $T = 5/g$, as a function of $g\tau$. (e)~Effective energy $\mathcal{E}(t)$ as a function of time for the viscous ($g\tau = 0.1$, green), intermediate ($g\tau = 1$, orange), and elastic ($g\tau = 100$, purple) regimes. (f)~Effective energy $\mathcal{E}(T)$ at $T = 5/g$ as a function of $g\tau$.}    \label{fig:uniaxial}
\end{figure*}

The torque balance equation for a slender beam then is given by (see~\cite{Landau1986}):
\begin{equation}
\frac{dT}{ds} = F \sin \theta(s,t),
\end{equation}
where $T$ is the torque applied to each beam element, $s$ is the arclength along the beam, $\theta$ is the angle between the tangential direction to the beam and the initial straight position, and $F$ is the compressive load (see Fig. \ref{fig:uniaxial}a) .  The torque is controlled by the bending of the beam:
\begin{equation}
    T(s_0, t) = I \int_{t_0}^t e^{-(t - t') / \tau} \, d \kappa(s_0, t'),
    \label{eq:torque}
\end{equation}
where $T$ is the torque, $I$ is the bending modulus of the beam --- the second moment of the cross-section with the elastic modulus of the material absorbed into it, see the SI --- the position $s_0$ is defined in the reference frame of the initially undeformed beam, and $\kappa$ is the local curvature. The contribution of curvature from earlier times to the current torque decays exponentially with a characteristic relaxation time $\tau$ (see the SI for the derivation). Note that growth does not appear explicitly in these equations. Instead, it enters through the compressive load $F$, which is determined by the constraint that the distance between the endpoints of the beam remains fixed while the arclength locally grows at the same rate.

Numerical solutions of these equations reveal that the dynamics is governed by a single dimensionless parameter: the product of the growth rate and the viscoelastic relaxation time, $g\tau$. The limit $g\tau \gg 1$ corresponds to elastic behavior, while $g\tau \ll 1$ corresponds to the viscous regime. For small buckling amplitudes, $g\tau$ plays no role: the beam shape is determined solely by its instantaneous length and takes the form of a sine wave (see the SI for the analytical solution in this limit). At larger buckling amplitudes, however, the sine wave is no longer a solution, and the behavior depends qualitatively on $g\tau$. In the elastic limit, the beam adopts a teardrop shape identical to a non-growing beam \cite{Levien:EECS-2008-103} but whose size increases over time, leading to a decrease in curvature. In the viscous limit, the torque is proportional to the rate of change of the curvature rather than to the curvature itself; consequently, instead of minimizing its curvature, the beam minimizes the rate at which curvature changes during the dynamics. As a result, the beam continues to coil, and the tip angle grows monotonically.

The most interesting scenario arises in the intermediate regime. A distinctive feature of this case is beam snapping (Fig.~\ref{fig:uniaxial}b): the beam rapidly transitions from one configuration to another. Physically, such a jump would be slowed by either inertia or friction with the surrounding medium, rather than occurring instantaneously as a direct transition between successive force-balance equilibria.

The snaps significantly accelerate the rate at which the beam coils, i.e.\ the rate at which the total number of turns increases (Fig.~\ref{fig:uniaxial}b). In the elastic limit, the beam does not coil at all, remaining in a teardrop shape; in the viscous limit, it coils slowly; and in the intermediate regime, coiling proceeds substantially faster owing to the snaps.

To understand the origin of these snaps, it is useful to consider the energy functional. Even though the regular energy is not conserved, still we can define the functional $\mathcal{E}_t$ at every moment of time such that the torque will be given by the variation of this functional  similar to \cite{goldstein2006} 

\begin{equation}
  T(s_0)_t = \frac{\delta \mathcal{E}[\theta, \theta', t]}{\delta \theta'(s_0)_t}.
\end{equation}

we assume here $t$ as a parameter, while $s$ is a variable. I.e., instead of one function $\theta(s,t)$ that depends on two variables we consider a set of functions $\theta_t(s)$ depending on only one variable $s$. Then we reproduce Eq.~\ref{eq:torque} for torque $T$ setting

\begin{eqnarray}
\mathcal{E} = \frac{I}{2 \tau}    \int_0^{L(t)} ds    \int_0^t d t'  \frac{
  e^{- (t - t')/\tau}}{1 - e^{-t/\tau} }   \left(
  \kappa(t) - \tilde{\kappa}(t,t')  \right)^2
\end{eqnarray}

where the arclength $s$ and the curvature $\kappa$ are both measured in the current configuration, $L(t) = L_0 e^{g t}$ is the current length of the beam, and each term $\tilde{\kappa}(t,t') =  (1 - e^{-t/\tau} )   \kappa(t') $  is the preferred curvature to which the beam would like to relax to minimize its energy (and each term contributes differently due to the viscoelastic relaxation). Written in these physical variables the growth rate does not appear explicitly: it enters only through the current length $L(t)$ and through the memory of the curvature.

The effective energy initially increases as the beam bends away from its straight configuration, but in both the viscous and elastic limits it quickly begins to decrease. In the intermediate regime, however, the energy rises again and subsequently exhibits a sawtooth pattern: it drops sharply each time the beam snaps and relaxes, then climbs back, repeating with every snap. Each snap thus corresponds to a transition from a higher-energy solution branch to a lower-energy one. We obtain energy and demonstrate how solution of the beam dynamics leads to snaps in the SI.

We see that in this model with one-dimensional growth the shape of the beam at each moment of time (with length measured in units of initial length and time in units of inverted growth rate) is controlled entirely by $g\tau$---not by the initial beam length or bending modulus---and that at intermediate values of $g\tau$ the behavior is qualitatively richer than in either the viscous or elastic limit. 

If we allow the beam to grow not only in length but also to increase in thickness, the beam's moment of inertia grows over time. Interestingly, however, this does not change the shape of the beam (as long as it remains thin); it only increases the torque and the compressive force in proportion to the growth rate (see the SI). 

The analysis in this section relied on the special geometry of a growing slender beam. In particular, we assumed that the stress-induced deformation of each material element remains small compared with the deformation produced directly by growth. For example, the beam was taken to thicken through growth alone, with any additional compression-induced change in thickness neglected. This separation makes the beam problem analytically tractable, but it does not apply to many other settings in which growth and mechanical deformation are intrinsically coupled.

The beam nevertheless reveals a broader principle: when the growth timescale becomes comparable to the viscoelastic relaxation time, that is, when $g\tau \sim 1$, the dynamics can become qualitatively distinct from both the viscous and elastic limits. To determine how general this behavior is, and under which geometric or mechanical constraints $g\tau \sim 1$ marks a critical threshold, we now develop a general theory of growing viscoelastic matter that treats growth-induced and stress-induced deformations on equal footing.

\section{Model of growing viscoelastic matter}

We begin by formulating the elastic limit in a way that is compatible with stress-free growth. Uniform, unconstrained growth should generate no stress, and after such growth the material should retain the same elastic response as before. The stress must therefore depend only on the elastic part of the deformation, with the deformation due to growth factored out. Following Ref.~\cite{rodriguez:94}, we decompose the deformation gradient tensor
\[
F_{ij}(\mathbf{x},t,t')=\frac{\partial x_i(t)}{\partial x_j(t')}
\]
into growth and elastic contributions,
\[
\mathbf{F}=\mathbf{F}_g\mathbf{F}_e.
\]
For isotropic unconstrained growth,
\[
\mathbf{F}_g=e^{(g/D)t}\mathbf{I},
\]
where $\mathbf{I}$ is the identity tensor, $g$ is the growth rate, and $D$ is the number of spatial dimensions. The same construction can be extended to growth rates that vary in space and time, as detailed in the SI. The stress tensor $\boldsymbol{\sigma}$ is then taken to depend only on the elastic part of the deformation:

\begin{eqnarray}
\boldsymbol{\sigma} = \mu\, \mathbf{B}_e - p\, \mathbf{I},
\end{eqnarray}
where we introduce the elastic part of the left Cauchy--Green strain tensor $\mathbf{B}_e(t, t') = \mathbf{F}_e(t, t')\, \mathbf{F}_e^T(t, t')$ in the  same way as the full strain tensor  $\mathbf{B} = \mathbf{F}\, \mathbf{F}^T$
in the neo-Hookean elasticity \cite{mooney1940,rivlin1948} (that is a generalization of standard Hooke's law allowing for big deformations and rotationally invariant; while it seems as the additional quadratic term in comparison with Hookean law, actually the elastic energy correspondent to it is also quadratic - see the SI). The pressure $p$ is determined by the compressibility properties of the material. By construction, the stress in this model is independent of uniform, unconstrained growth.

To incorporate viscoelasticity into this model, we allow the stress to be gradually forgotten through exponential decay. A standard way to achieve this is via the Lodge equation \cite{Larson2013}:

\begin{equation}
\boldsymbol{\tilde{\sigma}} = -\mu \int_{t_0}^t dt'\, e^{-(t - t') / \tau}\, d\mathbf{B}(t, t'),
\label{eq:lodge}
\end{equation}

where $t_0$ is the time at which the material was stress-free, $\boldsymbol{\tilde{\sigma}} = \boldsymbol{\sigma} + p\,\mathbf{I}$ is the stress tensor with the pressure $p$ subtracted, and $\tau$ is the characteristic relaxation time. The Lodge equation is originally written for the non-growing material, so the total strain tensor $\mathbf{B}$ contributes (the growth is zero so elastic part is the whole tensor). Taking a time derivative on both sides of Eq.~\ref{eq:lodge} one can obtain the differential form of this equation, or the upper-convected Maxwell model, which relates the stress tensor to the velocity field $\mathbf{v}$ (see the SI for the derivation):
\begin{equation}
\overset{\nabla}{\boldsymbol{\tilde{\sigma}}} = \mu \left(\boldsymbol{\nabla}\mathbf{v} + \boldsymbol{\nabla}\mathbf{v}^T\right) - \boldsymbol{\tilde{\sigma}} / \tau,
\end{equation}
where $\overset{\nabla}{\boldsymbol{\tilde{\sigma}}}$ denotes the upper-convected derivative, defined as
\begin{equation}
\overset{\nabla}{\boldsymbol{\tilde{\sigma}}} = \frac{\partial \boldsymbol{\tilde{\sigma}}}{\partial t} + \mathbf{v} \cdot \boldsymbol{\nabla} \boldsymbol{\tilde{\sigma}} - \boldsymbol{\nabla}\mathbf{v}\,\boldsymbol{\tilde{\sigma}} - \boldsymbol{\tilde{\sigma}}\,\boldsymbol{\nabla}\mathbf{v}^T.
\end{equation}
For an incompressible material, the pressure is chosen such that the velocity satisfies the continuity equation:
\begin{equation}
\boldsymbol{\nabla} \cdot \mathbf{v} = 0.
\end{equation}

To construct a model for growing viscoelastic matter, we follow the same strategy: we start from the growing neo-Hookean elastic solid and introduce stress relaxation via the Lodge equation,
\begin{equation}
\boldsymbol{\tilde{\sigma}} = -\mu \int_{t_0}^t e^{-(t - t') / \tau}\, d\mathbf{B}_e(t, t'),
\label{eq:lodge_grow}
\end{equation}
which is different from Eq.~\ref{eq:lodge} by having only the elastic part of the deformation. From this expression, we derive 
(see the SI) the upper-convected Maxwell model for a growing viscoelastic material:
\begin{widetext}
\begin{eqnarray}
\frac{\partial \boldsymbol{\tilde{\sigma}}}{\partial t} =
-\frac{1}{\tau}\,\boldsymbol{\tilde{\sigma}} +
\left(\boldsymbol{\nabla}\mathbf{v} - \frac{g}{D}\,\mathbf{I}\right) \boldsymbol{\tilde{\sigma}} +
\boldsymbol{\tilde{\sigma}} \left(\boldsymbol{\nabla}\mathbf{v}^T - \frac{g}{D}\,\mathbf{I}\right) +
\mu \left(\boldsymbol{\nabla}\mathbf{v} + \boldsymbol{\nabla}\mathbf{v}^T - \frac{2g}{D}\,\mathbf{I}\right) -
\mathbf{v} \cdot \boldsymbol{\nabla}\,\boldsymbol{\tilde{\sigma}}
\label{eq:const-viscoel-grow}
\end{eqnarray}
\end{widetext}
and the continuity equation becomes
\begin{equation}
\boldsymbol{\nabla} \cdot \mathbf{v} = g.
\end{equation}

We observe that every occurrence of $\boldsymbol{\nabla}\mathbf{v}$ in the standard non-growing model is replaced by $\boldsymbol{\nabla}\mathbf{v} - (g/D)\,\mathbf{I}$, ensuring that this term vanishes under uniform, unconstrained growth and remains small when the deformation deviates only slightly from the unconstrained growth. This structure resembles the one obtained in Ref.~\cite{zieger2026} with the difference in the relaxation: we use the linear Maxwell (Lodge) form, in which the stress decays at a fixed rate $1/\tau$, whereas Ref.~\cite{zieger2026} uses a relaxation term quadratic in $\mathbf{B}_e$, so that the relaxation rate itself grows with the accumulated elastic strain. 

When growth occurs much more slowly than the relaxation time, i.e.\ $g\tau \ll 1$, the system approaches the viscous limit that can be modeled by the unmodified Stokes equation and the modified matter conservation law as for example in Ref.~\cite{Giometto2018}. However, when this dimensionless parameter becomes of order unity, elastic effects can no longer be neglected and the full viscoelastic model must be used.

\begin{figure}
    \centering
    \includegraphics[width=0.95\linewidth]{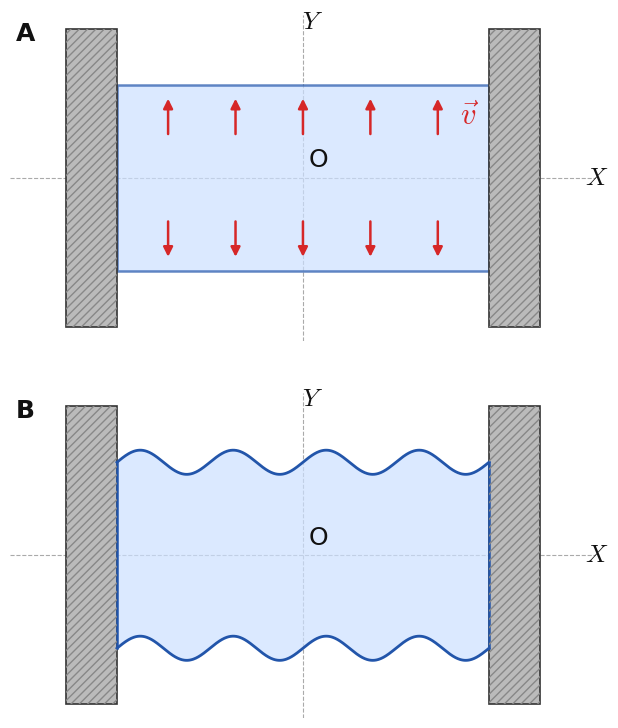}
    \caption{A. Cartoon of the growing material in a 2D channel. Walls constrain the growth in one direction only. B. Perturbation on (A).}
    \label{fig:channel}
\end{figure}

As $g\tau$ passes through a critical value of order unity, a bifurcation may occur. To see this, we consider perhaps the simplest possible constraint on growing matter: growth in a two-dimensional channel with frictionless walls (see Fig.~\ref{fig:channel}). In this geometry, growth in the $x$-direction is suppressed ($v_x = 0$), while growth in the $y$-direction is free. Solving Eq.~\ref{eq:const-viscoel-grow} in this setting yields $v_y = g\,y$ and
\begin{equation}
\sigma_{xx} =
-\mu\, g\tau \left[
\frac{1 - e^{-(1 + g\tau)\,t/\tau}}{1 + g\tau}
+
\frac{1 - e^{-(1 - g\tau)\,t/\tau}}{1 - g\tau}
\right].
\end{equation}
For $g\tau < 1$, a stationary solution exists:
\begin{equation}
\sigma_{xx} =
-\mu\, g\tau \left[
\frac{1}{1 + g\tau}
+
\frac{1}{1 - g\tau}
\right],
\end{equation}
which diverges as $g\tau \to 1$. For $g\tau \geq 1$, no stationary solution exists: the stress grows exponentially in time. In practice, the stress would eventually become large enough for the assumptions underlying the model to break down.

The system thus undergoes a phase transition at $g\tau = 1$: below this threshold, the stress builds up gradually toward a stationary value; above it, the stress grows exponentially without bound. As before, this behavior is independent of the Young modulus $\mu$ or the channel width---the only parameter that matters is $g\tau$.

In the SI, we build on this geometry to explore the effects of $g\tau$ in two more biologically relevant scenarios: growth in a porous medium (with a single pore modeled as a channel with no-slip walls) and wrinkling instabilities of a thin growing layer (mimicking a wrinkling biofilm; see the cartoon in Fig.~\ref{fig:channel}B). For growth in a porous medium, we show that the permeability increases in the viscoelastic case by a factor of $1 + 2.5 g \tau$ for small $g \tau$. Notably, a growing viscous fluid and a non-growing viscoelastic fluid have the same permeability as a standard non-growing viscous fluid: only the combination of viscoelasticity and growth alters the permeability. The derivation and a qualitative explanation of the sign of this effect are provided in the SI.

For the wrinkling instabilities, we find that the wrinkle amplitude depends on $g\tau$ in a non-monotonic fashion, reaching a maximum at intermediate values. This maximum can exceed the purely viscous and purely elastic values by several times. The derivation is given in the SI.

\section{Discussion}

In this work, we constructed a model for viscoelastic growing matter, which recovers the correct limits in purely elastic, purely viscous, and non-growing cases. We found a single dimensionless control parameter, $g \tau$, the product of growth rate $g$ and viscoelastic relaxation time $\tau$. This parameter determines whether the material behaves more like an elastic solid or a viscous fluid. When $g \tau = 0$, stress relaxes faster than builds up due to the growth, and the system behaves as a viscous fluid. In contrast, when $g \tau \gg 1$, growth is fast compared to relaxation and the behavior approaches the elastic limit. The viscoelastic behavior in the intermediate regime $g \tau \approx 1$ appears to be qualitatively different from both viscous and elastic limits.

In natural systems, the values of $g \tau$ can vary significantly in the same material. This variability arises for at least two reasons. First, growth is often non-homogeneous and depends on external factors such as nutrient concentration, leading different regions of the same tissue or biofilm to operate in different mechanical regimes. Second, real biological and polymeric materials are more complex than the single-component Maxwell model considered here and often possess multiple relaxation time scales. In such cases, the total stress is a sum of contributions from several viscoelastic modes,

\begin{equation}
\boldsymbol{\tilde{\sigma}}_{\rm tot} = \sum_i \boldsymbol{\tilde{\sigma}}_{i} ,
\end{equation}

where

\begin{equation}
\boldsymbol{\tilde{\sigma}}_i = - \mu_i \int_{t_0}^t dt' e^{-(t - t') / \tau_i} d \mathbf{B}(t, t'),
\label{eq:lodge-modes}
\end{equation}

For example, in biofilms, the relaxation times of different components span a wide range—from seconds to several minutes \cite{peterson13} \cite{klapper02}. The slowest viscoelastic mode has been measured at approximately $\tau = 18$ minutes, a universal time across multiple species \cite{shaw04}. Combined with typical biofilm growth rates—which are slower than those of well-mixed cultures due to the metabolic cost of producing extracellular matrix—the resulting dimensionless parameter $g \tau$ can reach values on the order of $0.1$. Although this is still smaller than unity, the effects are nonetheless significant in the scenarios discussed above. In the case of flow through a channel, the permeability coefficient is modified by approximately 25 \%, while for the wrinkling beam, the maximal amplitude increases by roughly 50\% for the parameters used in Fig.~\ref{fig:curves}. 

For other biological tissues, the situation is more complex. They are often modeled using an “infinite” viscoelastic component, i.e., with $\tau = \infty$, representing a purely elastic contribution. However, strictly speaking, no component is truly elastic: over sufficiently long time scales, even the collagen network that provides structural integrity can remodel and relax accumulated stress in a turnover process. This relaxation occurs both during growth and in the absence of external loading, but such long relaxation times—ranging from months to years—are extremely difficult to measure experimentally. Nevertheless, for sufficiently slow growth rates, the product $g \tau$ in tissues may approach values of order 1, placing them squarely in the viscoelastic regime rather than in the elastic or viscous limits.

The geometries considered here are deliberately simplified model systems. They do not account for the full three-dimensional structure of biofilms or tissues, the spatially heterogeneous growth driven by nutrient gradients, or the elasticity of the surrounding porous matrix in which biofilms develop. Nevertheless, our results demonstrate that even in such idealized settings, viscoelastic growing matter exhibits qualitatively distinct behavior that cannot be captured by purely elastic or purely viscous models. This underscores the necessity of incorporating viscoelastic effects in more detailed, biologically realistic models of growth and morphogenesis.
Several extensions are left for future work. First, growth in a medium that creates bulk friction (e.g., from a non-growing extracellular matrix) requires the total stress to be balanced both by forces from the medium and by the deformation of the growing matter studied here. Second, non-uniform growth caused by nutrient depletion is already described by the general theory we have introduced — since we never assumed uniformity — but we did not explore any specific examples of such systems. Third, anisotropic growth poses an additional challenge: decomposing the full deformation gradient tensor into growth and elastic parts becomes ambiguous because these contributions do not commute in general. Addressing this will require either more sophisticated decomposition strategies or the use of our equations in differential form as a starting definition.

\begin{acknowledgments}
This work was supported by the Deutsche Forschungsgemeinschaft (DFG) within the SPP 2389 priority program ``Emergent Functions of Bacterial Multicellularity'' (Grant number: 503995533). OH acknowledges support by a Humboldt Professorship of the Alexander von Humboldt Foundation. We used Claude (Anthropic) to help with code for the numerical implementation,  check analytical derivations and their consistency with numerical solution, and to refine the style of our text. VS would like to thank Katja Taute for helpful comments and suggestions.
\end{acknowledgments}

\section*{Data Availability}

The code used to produce the numerical results and the figures of this work is published on GitHub, \url{https://github.com/valentinslepukhin/growing-viscoelastic-beam}.

\appendix

% --- Supplemental Information is typeset in a single column ---
\onecolumngrid

\section{General model for viscoelastic growing matter}

To present our self-consistent model of viscoelastic growing matter, we begin with a brief review of established approaches. Specifically, we summarize standard models for growing elastic matter, growing viscous fluids, and non-proliferating viscoelastic matter. We then discuss existing attempts to model proliferating viscoelastic matter and highlight their main limitations.

\subsection{Growing viscous fluid.}

A standard way to model a proliferating viscous fluid, used for example in \cite{Giometto2018}, is to modify the mass conservation law:

\begin{equation}
  \frac{\partial \rho}{\partial t}   = - \boldsymbol{\nabla} \cdot (\mathbf{v} \rho) + g \rho
\end{equation}

where $\rho$ denotes the cell density, $\mathbf{v}$ the velocity field, and $g$ the growth rate. This formulation reflects that matter is not conserved, but instead grows at rate $g$. In the incompressible limit with uniform cell density, the time derivative on the left-hand side vanishes, leading to

\begin{equation}
0 = - \boldsymbol{\nabla} \cdot \mathbf{v} + g
\label{eq:continuity-incompressible}
\end{equation}

Throughout the paper $g$ denotes this \emph{volumetric} growth rate, i.e.\ the trace of the growth velocity gradient, so that a material volume element grows as $e^{g t}$ regardless of geometry. How the total rate $g$ is distributed among the individual axes, however, depends on the geometry and on the constraints. For isotropic growth in $D$ dimensions each axis elongates at rate $g / D$, so that the growth part of the deformation gradient is $\mathbf{F}_g = e^{g t / D} \mathbf{I}$ (see Eq.~\ref{eq:F-growth-shear} below for the two-dimensional case). When the growth is directed, as in the beam geometries of the following sections, the whole of $g$ is carried by a single axis; the exponent $e^{g t}$ that then appears for that axis is the same $g$ as in Eq.~\ref{eq:continuity-incompressible}, only partitioned differently.

The stress is still governed by the standard Stokes equation for an incompressible fluid:

\begin{equation}
\boldsymbol{\sigma} = \eta \left( \boldsymbol{\nabla} \mathbf{v} + \boldsymbol{\nabla} \mathbf{v}^T \right) - p \mathbf{I},
\end{equation}

where $\eta$ is the viscosity and $p$ is a Lagrange multiplier enforcing the incompressibility constraint. Importantly, in the case of a growing fluid, $p$ does not correspond to the physical pressure (the trace of the stress tensor) but instead includes a contribution from the growth rate. This becomes clear in the case of uniform, unconstrained growth ($g = \mathrm{const}$), where the stress must vanish since growth is homogeneous. In that situation, one finds $ p = \frac{2}{D} g \eta$, where $D$ is the number of spatial dimensions. In this approach, this redefinition of $p$ does not play a physical role, since $p$ is just a Lagrange multiplier, and physically we measure the stress tensor. However, as we go to the elastic theory, we will see that the effect of growth is more complicated there. 

\subsection{Growing elastic material}

Throughout, we typeset tensors and vectors in boldface and their components in plain italic, so that $\mathbf{F}$ is a tensor and $F_{ij}$ one of its components. Latin symbols are set upright bold ($\mathbf{F}$, $\mathbf{B}$, $\mathbf{v}$, $\mathbf{I}$) and Greek ones italic bold ($\boldsymbol{\sigma}$, $\boldsymbol{\tilde{\sigma}}$, $\boldsymbol{\omega}$); subscripts always sit outside the boldface group, as in $\mathbf{F}_e$ and $\boldsymbol{\tilde{\sigma}}_i$. The gradient operator is likewise bold, $\boldsymbol{\nabla}$.

\begin{figure}
    \centering
    \includegraphics[width=0.9\linewidth]{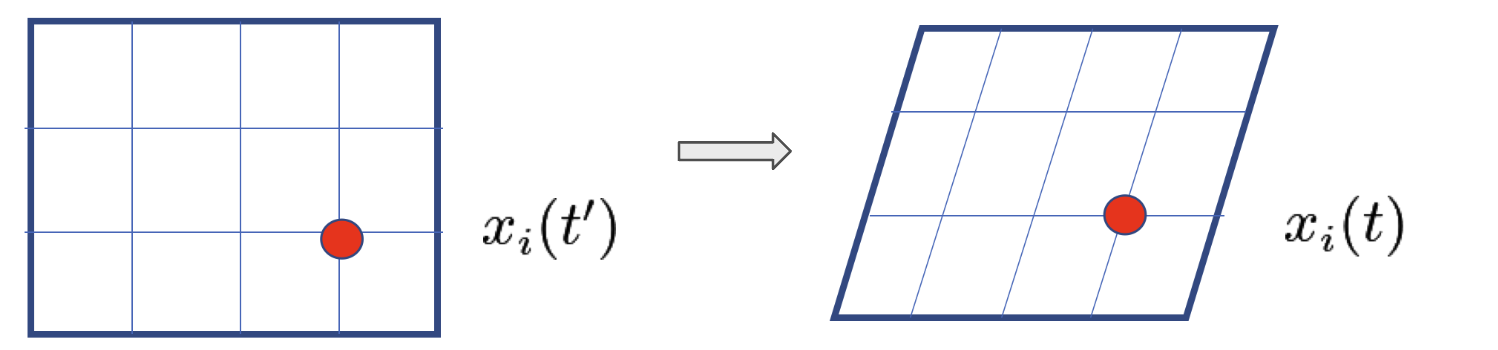}
    \caption{The illustration of the deformation of the object from moment of time $t'$ to the moment of time $t$.}
    \label{fig:def-tensor}
\end{figure}

In standard elasticity theory, the deformation of a body is characterized by the deformation gradient tensor $\mathbf{F}$. This tensor describes how each material point is displaced at time $t$ relative to its reference configuration at an earlier time $t'$ (see Fig.~\ref{fig:def-tensor}).

\begin{eqnarray}
    F_{i j}(x(t'), t ,t') = \frac{\partial x_i(t)}{\partial x_j(t')}
    \label{eq:def-tensor}
\end{eqnarray}

This tensor has a convenient property: it allows the composition of successive deformations,

\begin{equation}
\mathbf{F}(t_3, t_1) = \mathbf{F}(t_3, t_2) \mathbf{F}(t_2, t_1) .
\label{eq:superpos}
\end{equation}

In the limit of small deformations (the Hookean regime), all displacements are small and $\mathbf{F}$ is close to the identity matrix. In this case,

\begin{equation}
F_{ij}(x(t'), t ,t')= \delta_{ij} + \frac{\partial \Delta x_i}{\partial x_j}
= \delta_{ij} + \partial_j u_i ,
\end{equation}

where $\mathbf{u}$ denotes the displacement field.

Hooke’s law for an isotropic material states that the stress is proportional to the displacement gradient:

\begin{equation}
\boldsymbol{\sigma} = \mu \left( \boldsymbol{\nabla} \mathbf{u} + (\boldsymbol{\nabla} \mathbf{u})^T \right) + \lambda  (\boldsymbol{\nabla} \cdot \mathbf{u})  \mathbf{I},
\end{equation}

where $\mu$ (the shear modulus) characterizes the response to shear deformations, and $\lambda$ (the first Lamé parameter) controls the response to isotropic compression.

To ensure invariance under stress-free deformations (such as rigid body rotations, which yield a non-identity $F$), one often employs the neo-Hookean model. In this framework, the stress is expressed not in terms of $\mathbf{F}$ directly, but through the rotation-invariant left Cauchy–Green strain tensor $\mathbf{B}$, defined as

\begin{equation}
B_{ij}(t,t') = F_{ik}(t,t') F_{jk}(t,t'),
\end{equation}

or, equivalently in matrix form,

\begin{equation}
\mathbf{B} = \mathbf{F}\mathbf{F}^T .
\end{equation}

In the incompressible case, which is the main focus of our study, the stress is defined to be proportional to the left Cauchy–Green tensor $\mathbf{B}$:

\begin{equation}
\boldsymbol{\sigma} = \mu  \mathbf{B} - p \mathbf{I},
\label{eq:neohook}
\end{equation}

where $p$ again acts as a Lagrange multiplier enforcing incompressibility. It is straightforward to verify that, in the limit of small deformations, the neo-Hookean model reduces to the classical Hookean form.

A straightforward approach to incorporate growth into the model would be to follow a procedure analogous to the viscous case. Specifically, one could modify the incompressibility constraint to account for growth, which would in turn adjust the pressure. For instance, one might enforce

\begin{eqnarray}
    \det F(t_1, t_2) = e^{\int_{t_1}^{t_2} g(t) dt } 
    \label{eq:neohook-incomp}
\end{eqnarray}

\begin{figure}
    \centering
    \includegraphics[width=0.9\linewidth]{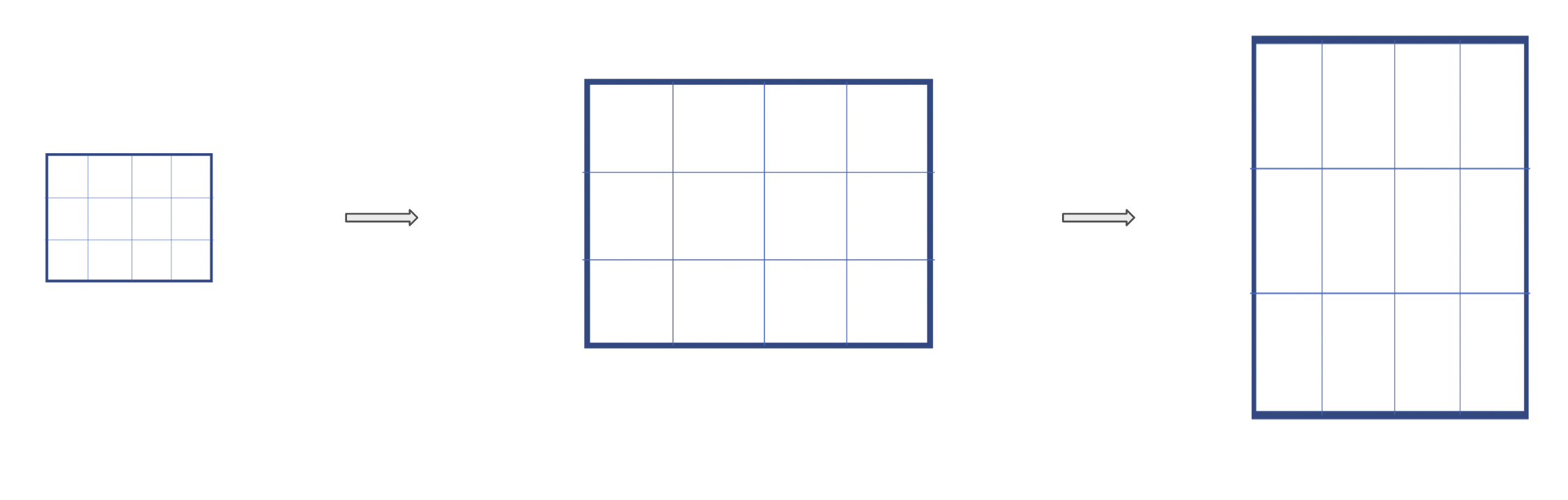}
    \caption{Illustration for the uniform growth followed by an isochoric stretch of ratio $\gamma$.}
    \label{fig:growth-shear}
\end{figure}

To illustrate the problem, consider a 2D object that first grows uniformly and isotropically at a constant rate $g$ without any applied stress, and is then subjected to an isochoric (area-preserving) stretch of ratio $\gamma$ (see Fig.~\ref{fig:growth-shear}). Note that $\gamma$ is a stretch \emph{ratio}, not a strain: the undeformed state is $\gamma = 1$, and the object is stretched along $x$ and compressed along $y$ by the same factor. This is the principal-axes form of a pure shear; we call it an isochoric stretch to avoid confusion with simple shear, whose deformation gradient is off-diagonal.

In this case, the total deformation gradient tensor can be written as the product of the growth and stretch components:

\begin{equation}
    \mathbf{F} = \mathbf{F}_s \mathbf{F}_g =
     \begin{pmatrix}
        \gamma & 0
        \\
        0 & \frac{1}{\gamma} 
    \end{pmatrix}
    \begin{pmatrix}
        e^{g t / 2} & 0
        \\
        0 & e^{g t / 2}
    \end{pmatrix}
    \label{eq:F-growth-shear}
\end{equation}

    Then the stress in the assumption of neo Hookean elasticity Eq.~\ref{eq:neohook} is

\begin{equation}
     \boldsymbol{\sigma} = \mu 
     e^{ g t} \begin{pmatrix}
         \gamma^2  & 0 
        \\
        0 & \frac{1}{\gamma^2} 
     \end{pmatrix}- p \mathbf{I}
\end{equation}

   where $p$ is determined by the incompressibility condition Eq. ~\ref{eq:neohook-incomp}. The difference between normal stresses $\sigma_{xx} - \sigma_{yy} $ is a physically measured quantity and is pressure independent

   \begin{equation}
    \sigma_{xx} - \sigma_{yy} = \mu 
     e^{ g t} \left( 
         \gamma^2  - \frac{1}{\gamma^2} \right)   
         \label{eq:naive-neohook}
\end{equation}
 
     We observe that, in this model, the stress depends on the amount of uniform growth that occurred prior to the stretch. This is unphysical: after uniform, unconstrained growth, a material element should have the same mechanical properties as an element of the same size before growth. However, according to Eq.~\ref{eq:naive-neohook}, sufficiently large growth can generate a substantial stress even under a stretch ratio close to unity. This motivates the need to explicitly incorporate in the model the principle that uniform growth alone does not produce stress.

This approach was introduced in~\cite{rodriguez:94}, where the deformation gradient tensor is explicitly decomposed into a growth component and an elastic component: $\mathbf{F} = \mathbf{F}_g \mathbf{F}_e$. For isotropic growth, $\mathbf{F}_g$ is proportional to the identity matrix, so it commutes with any other matrix and the order in the product does not matter. In this framework, the stress depends only on the elastic part of the deformation:

   \begin{eqnarray}
    \boldsymbol{\sigma} = \mu \mathbf{F}_e\mathbf{F}_e^T  - p \mathbf{I}
\end{eqnarray}

where the pressure $p$ is still determined by the incompressibility constraint (Eq.~\ref{eq:neohook-incomp}). In this model, the stress is, by construction, explicitly independent of uniform, unconstrained growth. For the example considered above, we then obtain

\begin{equation}
    \sigma_{xx} - \sigma_{yy} = \mu  \left( 
         \gamma^2  - \frac{1}{\gamma^2} \right)   
\end{equation}

which is entirely independent of the amount of growth.

\subsection{Viscoelastic matter}

A viscoelastic material can "remember" stress over short timescales (smaller than the viscoelastic relaxation time $\tau$) and gradually relax it over longer timescales. One way to describe this behavior is to start from the purely elastic case and introduce an exponential decay of stress. To do so, we first express the stress at a later time in terms of the stress at an earlier time, and then include a decay term.

We introduce the excess stress $\boldsymbol{\tilde{\sigma}}$, which is defined to be zero in the undeformed state by construction:

\begin{equation}
    \boldsymbol{\tilde{\sigma}}(t) = \mu \left( \mathbf{F}(t, t_0) \mathbf{F^T}(t, t_0)  -  \mathbf{I} \right) 
\end{equation}

where $t_0$ is the reference time at which the material is stress-free.

The total stress is given then by the equation

 \begin{eqnarray}
    \boldsymbol{\sigma} =  \boldsymbol{\tilde{\sigma}}  - (p - \mu) \mathbf{I}
\end{eqnarray} 
    
We can express the excess stress at the next time step as

\begin{equation}
    \boldsymbol{\tilde{\sigma}}( t + dt) = \mu \mathbf{F}(  t + dt, t_0) \mathbf{F^T}( t, t + dt, t_0)  -  \mu \mathbf{I}
\end{equation}

Using the composition property of the deformation gradient (Eq.~\ref{eq:superpos}), we can write

\begin{equation}
    \boldsymbol{\tilde{\sigma}}(t + dt) = \mu  \mathbf{F}( t + dt, t) \mathbf{F}(t , t_0) \mathbf{F^T}(t , t_0)  \mathbf{F^T}( t + dt, t)  - \mu\mathbf{I} 
\end{equation}

which naturally leads to an expression relating the stress at the next moment to the stress at the previous moment and the deformation between these two times:

\begin{equation}
    \boldsymbol{\tilde{\sigma}}(t + dt) =  \mathbf{F}( t + dt, t)  ( \boldsymbol{\tilde{\sigma}}(t ) + \mu \mathbf{I})   \mathbf{F^T}( t + dt, t)   - \mu \mathbf{I} 
    \label{eq:sigma-ev}
\end{equation}

To incorporate stress relaxation, we introduce an exponential decay on the right-hand side of the evolution equation, giving:

\begin{equation}
    \boldsymbol{\tilde{\sigma}}(t + dt) =  \mathbf{F}( t + dt, t)  ( \boldsymbol{\tilde{\sigma}}(t ) e^{-dt / \tau} + \mu \mathbf{I})   \mathbf{F^T}( t + dt, t)   - \mu \mathbf{I} 
    \label{eq:sigma-ev-visc}
\end{equation}

where $\tau$ is the viscoelastic relaxation time. This formulation ensures that the material gradually “forgets” its previous stress over timescales larger than $\tau$. This equation is equivalent to the integral Lodge equation for viscoelastic matter~\cite{Larson2013}

\begin{equation}
  \boldsymbol{\tilde{\sigma}} = -\mu \int_{t_0}^t e^{-(t - t') / \tau} d \mathbf{B}(t, t'),
  \label{eq:lodge-si}
\end{equation}
where $t_0$ is a moment of time when the object was unstressed. This can be observed using the definition of the tensors $\mathbf{F}$ and $\mathbf{B}$, composition property,  and expanding for small time step $dt$.

Note that Eq.~\ref{eq:sigma-ev-visc} no longer depends explicitly on times earlier than $t$. However, it still carries an implicit dependence on the initial configuration, since all coordinates are expressed in the reference frame of the undeformed material. Making this dependence explicit, let $X$ denote the position in the initial, stress-free configuration, and $x(X, t)$ the current position of this point after deformation. Then the excess stress can be written as

    \begin{equation}
    \boldsymbol{\tilde{\sigma}}(x(X, t+dt), t + dt) =  \mathbf{F}( t + dt, t)  ( \boldsymbol{\tilde{\sigma}}(x(X, t) , t ) e^{-dt / \tau}  + \mu \mathbf{I}  ) \mathbf{F^T}( t + dt, t)  - \mu \mathbf{I} 
\end{equation}

The displacement of a material point over a small time interval $dt$ is determined by the velocity field:

\begin{eqnarray}
    x(X, t+dt) = x(X,t) + v(x(X,t),t) dt
\end{eqnarray}

Using this relation together with the definition of the deformation gradient tensor $\mathbf{F}$ (Eq.~\ref{eq:def-tensor}), we can write the stress evolution in terms of quantities evaluated at time $t$ and position $x(X,t)$:

    \begin{equation}
    \boldsymbol{\tilde{\sigma}} + \frac{\partial \boldsymbol{\tilde{\sigma}}
}{\partial t} dt  + \mathbf{v} \cdot \boldsymbol{\nabla}  \boldsymbol{\tilde{\sigma}}dt  = (\mathbf{I} + \boldsymbol{\nabla}\mathbf{v} dt )  (\boldsymbol{\tilde{\sigma}} e^{-dt / \tau}  + \mu \mathbf{I}  ) (\mathbf{I} + \boldsymbol{\nabla}\mathbf{v}^T dt )   - \mu \mathbf{I}  
\end{equation}

 Expanding it for small $dt$ and rearranging terms, we obtain a differential equation describing the time evolution of the excess stress:

\begin{equation}
      \hat{D}   \boldsymbol{\tilde{\sigma}}  =  \mu (\boldsymbol{\nabla}\mathbf{v} +\boldsymbol{\nabla}\mathbf{v}^T )  - \boldsymbol{\tilde{\sigma}} / \tau, 
\end{equation}

where $\hat{D}$ is the convective derivative, defined as

\begin{equation}
     \hat{D}   \boldsymbol{\tilde{\sigma}} =  \frac{\partial \boldsymbol{\tilde{\sigma}}
}{\partial t}   + \mathbf{v} \cdot \boldsymbol{\nabla}  \boldsymbol{\tilde{\sigma}}  -   \boldsymbol{\nabla}\mathbf{v} \boldsymbol{\tilde{\sigma}}   - \boldsymbol{\tilde{\sigma}} \boldsymbol{\nabla}\mathbf{v}^T 
\label{eq:maxwell}
\end{equation}

Equation \ref{eq:maxwell} is equivalent to the upper-convected Maxwell model commonly used in viscoelastic fluid theory. The convective derivative $\hat{D}$ ensures that the stress is properly “transported” and rotated with the flow, while the term $-\boldsymbol{\tilde{\sigma}}/\tau$ accounts for stress relaxation. In the limit $\tau \to \infty$, the material behaves as a purely elastic solid, whereas for finite $\tau$ it exhibits viscoelastic behavior, transitioning between solid-like and fluid-like responses. In the opposite limit, $\tau \to 0$, the material behaves as a viscous fluid with an effective viscosity $\eta = \mu  \tau$. This formulation also coincides with the Oldroyd-B model in the limit of negligible solvent viscosity. 

The total stress is then given by

\begin{eqnarray}
    \boldsymbol{\sigma} =  \boldsymbol{\tilde{\sigma}}  - (p - \mu) \mathbf{I}
\end{eqnarray}  

where $p$ enforces the incompressibility condition. In the purely elastic case, this condition was written as $\det \mathbf{F} = 1$. Expressing it in terms of the velocity field at the current time leads to the standard continuity equation:

\begin{equation}
\boldsymbol{\nabla} \cdot \mathbf{v}  = 0
\end{equation}

\subsection{Growing viscoelastic material}

Models that incorporate both viscoelasticity and growth have been suggested, for example, in \cite{Chen2025Chirality, Parmar2025ProliferatingNematic}, where chiral viscoelastic growing media were considered. In these approaches, growth is introduced by modifying the matter conservation equation, in analogy with the purely viscous case:

\begin{equation}
\boldsymbol{\nabla} \cdot \mathbf{v} = g,
\end{equation}

while the stress evolution equation remains unchanged. However, this can lead to contradictions in the elastic limit. Specifically, taking the limit $\tau \to \infty$ in the viscoelastic theory should recover the purely elastic case, with stress determined solely by the elastic deformation and a modified incompressibility constraint. This corresponds to the “naive” approach described in the previous section, which produces an unphysical dependence of stress on prior growth (Eq.~\ref{eq:naive-neohook}).

The simplified growth-modified viscoelastic model can be used when the deformation due to growth over a timescale comparable to the relaxation time $\tau$ is small, which is controlled by the dimensionless parameter $g \tau$. However, when this quantity is not small, the effect of growth on stress cannot be neglected, and the model must be modified to account for it explicitly.

The most straightforward way to obtain a self-consistent viscoelastic model with growth is to start from the full elastic model with growth—where the total deformation tensor $\mathbf{F}$ is replaced by the elastic deformation tensor $\mathbf{F}_e$—and then apply the same procedure used previously to derive the viscoelastic version. Namely, we start with

\begin{equation}
    \boldsymbol{\tilde{\sigma}}(t) = \mu \left( \mathbf{F}_e(t, t_0) \mathbf{F}_e^T(t, t_0)  -  \mathbf{I} \right) 
\end{equation}

where $t_0$ is the reference time at which the material is stress-free.

We can express the excess stress at the next time step as

\begin{equation}
    \boldsymbol{\tilde{\sigma}}( t + dt) = \mu \mathbf{F}_e(  t + dt, t_0) \mathbf{F}_e^T( t, t + dt, t_0)  -  \mu \mathbf{I}
\end{equation}

The composition property is true for the elastic part of the deformation tensor as well, so

\begin{equation}
    \boldsymbol{\tilde{\sigma}}(t + dt) = \mu  \mathbf{F}_e( t + dt, t) \mathbf{F}_e(t , t_0) \mathbf{F}_e^T(t , t_0)  \mathbf{F}_e^T( t + dt, t)  - \mu\mathbf{I} 
\end{equation}

which naturally leads to an expression relating the stress at the next moment to the stress at the previous moment and the deformation between these two times:

\begin{equation}
    \boldsymbol{\tilde{\sigma}}(t + dt) =  \mathbf{F}_e( t + dt, t)  ( \boldsymbol{\tilde{\sigma}}(t ) + \mu \mathbf{I})   \mathbf{F}_e^T( t + dt, t)   - \mu \mathbf{I} 
\end{equation}

To incorporate stress relaxation, we introduce an exponential decay on the right-hand side of the evolution equation, giving:

\begin{equation}
    \boldsymbol{\tilde{\sigma}}(t + dt) =  \mathbf{F}_e( t + dt, t)  ( \boldsymbol{\tilde{\sigma}}(t ) e^{-dt / \tau} + \mu \mathbf{I})   \mathbf{F}_e^T( t + dt, t)   - \mu \mathbf{I} 
\end{equation}

where $\tau$ is the viscoelastic relaxation time.

Let $X$ denote the position in the initial, stress-free configuration, and $x(X, t)$ the current position of this point after deformation. Then the excess stress can be written as

    \begin{equation}
    \boldsymbol{\tilde{\sigma}}(x(X, t+dt), t + dt) =  \mathbf{F}_e( t + dt, t)  ( \boldsymbol{\tilde{\sigma}}(x(X, t) , t ) e^{-dt / \tau}  + \mu \mathbf{I}  ) \mathbf{F}_e^T( t + dt, t)  - \mu \mathbf{I} 
\end{equation}

The displacement of a material point over a small time interval $dt$ is determined by the velocity field:

\begin{eqnarray}
    x(X, t+dt) = x(X,t) + v(x(X,t),t) dt
\end{eqnarray}

Using this relation together with the definition of the elastic part of the deformation gradient tensor $\mathbf{F}$ , we can write the stress evolution in terms of quantities evaluated at time $t$ and position $x(X,t)$:

    \begin{equation}
    \boldsymbol{\tilde{\sigma}} + \frac{\partial \boldsymbol{\tilde{\sigma}}
}{\partial t} dt  + \mathbf{v} \cdot \boldsymbol{\nabla}  \boldsymbol{\tilde{\sigma}}dt  = e^{- \frac{g}{D} dt} (\mathbf{I} + \boldsymbol{\nabla}\mathbf{v} dt )  (\boldsymbol{\tilde{\sigma}} e^{-dt / \tau}  + \mu \mathbf{I}  ) (\mathbf{I} + \boldsymbol{\nabla}\mathbf{v}^T dt ) e^{-\frac{g}{D} dt}   - \mu \mathbf{I}  
\end{equation}

 Expanding it for small $dt$ and rearranging terms, we obtain a differential equation describing the time evolution of the excess stress:

\begin{eqnarray}
    \frac{\partial \boldsymbol{\tilde{\sigma}} }{\partial t}    =
- \frac{1}{\tau} \boldsymbol{\tilde{\sigma}}  +
\left( \boldsymbol{\nabla}\mathbf{v} - \frac{g}{D} \mathbf{I} \right) \boldsymbol{\tilde{\sigma}}+
  \boldsymbol{\tilde{\sigma}} \left( \boldsymbol{\nabla}\mathbf{v}^T - \frac{g}{D}
  \mathbf{I} \right) +
\mu \left( \boldsymbol{\nabla}\mathbf{v}+ \boldsymbol{\nabla}\mathbf{v}^T - \frac{2 g}{D} \mathbf{I} \right)
- \mathbf{v} \cdot \boldsymbol{\nabla}  \boldsymbol{\tilde{\sigma}}
\label{eq:const-viscoel-grow-app}
\end{eqnarray}

Note that here we did not assume $g$ to be constant in space, and can have $g(x)$ everywhere. Moreover, adding time dependence will only change the growth tensor from $e^{g t}$ to $e^{\int g(t) dt } $. 

Anisotropy, however, would modify the derivation, since we explicitly assume that growth and elastic deformation tensors commute. The anisotropic growth tensor does not have this property in general. Since the definition of the elastic tensor then becomes ambiguous, we may instead use Eq.~\ref{eq:const-viscoel-grow-app} with $\frac{g}{D} \mathbf{I}$ replaced by the growth rate tensor, as the definition of the model for anisotropically growing viscoelastic matter. 

\subsection{Matter growing within a substrate}

In the previous sections, we separately considered growth, free of stress, and the externally imposed stress. However, in many cases the growth happens in a substrate that is itself not growing. In this situation, the stress is imposed to the growing material from the substrate, and the force balance requires $\boldsymbol{\sigma}_{\rm material} = - \boldsymbol{\sigma}_{\rm substrate}$ .  

The deformation of the substrate can depend on the stress and the deformation of the material in different ways. In the limit of no mechanical interaction, we have undeformed substrate and freely growing matter. In the limit when the matter sticks to the substrate, the deformation of the substrate is equal to the deformation of the matter. Then, if the substrate is viscoelastic, its deformation is controlled by

\begin{eqnarray}
    \frac{\partial \boldsymbol{\tilde{\sigma}} }{\partial t}    =
- \frac{1}{\tau} \boldsymbol{\tilde{\sigma}}  +
\left( \boldsymbol{\nabla}\mathbf{v} \right) \boldsymbol{\tilde{\sigma}}+
  \boldsymbol{\tilde{\sigma}} \left( \boldsymbol{\nabla}\mathbf{v}^T \right) +
\mu \left( \boldsymbol{\nabla}\mathbf{v}+ \boldsymbol{\nabla}\mathbf{v}^T \right)
- \mathbf{v} \cdot \boldsymbol{\nabla}  \boldsymbol{\tilde{\sigma}}
\label{eq:const-viscoel-substrate}
\end{eqnarray}

with the velocity field coming from the growing matter and subject to the continuity equation for the growing matter

\begin{equation}
0 = - \boldsymbol{\nabla} \cdot \mathbf{v} + g
\end{equation}

    In such a setting, the stress of the viscoelastic non-growing substrate, subjected to the growing viscoelastic matter, is described exactly by the equations of \cite{Chen2025Chirality, Parmar2025ProliferatingNematic}.

In the intermediate regime, there can be a friction to the substrate, so in addition to the equation for the matter (with the velocity field and the stress of the matter) and for the substrate (with the velocity field of the substrate), we would have two force balance equations 

\begin{equation}
   \boldsymbol{\nabla} \cdot \boldsymbol{\sigma}_{\rm material} = - \boldsymbol{\nabla} \cdot \boldsymbol{\sigma}_{\rm substrate} = \Gamma (\mathbf{v}_{\rm material} - \mathbf{v}_{\rm substrate})
\end{equation}

 where $\Gamma$ is a friction coefficient between substrate and the growing material.

\subsection{Energy based formalism}

In a purely elastic, non growing case one may introduce the elastic energy. In incompressible limit, it is simply $\mathcal{E} = \frac{\mu}{2} \, {\rm tr} \mathbf{B} $, such that

\begin{equation}
    \boldsymbol{\sigma} = \frac{\delta \mathcal{E} }{\delta  \mathbf{F}}  \mathbf{F}^T  - p \mathbf{I}
    \label{eq:stress-energy}
\end{equation}
where pressure  $p$ is Lagrange multiplier ensuring incompressibility.

In a growing case, the energy is not conserved because of the explicit time dependence. However, the functional $\mathcal{E} = \frac{\mu}{2} \, {\rm tr} \mathbf{B}_e $  (the elastic energy at each moment of time) still leads to correct stress defined by Eq.~\ref{eq:stress-energy}. 

In a viscoelastic case, the energy is not conserved as well. However, we still have an explicitly time dependent functional $\mathcal{E}_t $ at each moment of time $t$, such that

\begin{equation}
    \boldsymbol{\sigma}_t = \frac{\delta \mathcal{E}_t }{\delta  \mathbf{F}_t}  \mathbf{F}_t^T 
\end{equation}

Here the subscript $t$ denotes that energy, stress and deformation gradient tensor are taken at the time $t$, so $t$ is a parameter here rather than variable. 

To find the form of this functional, we rewrite the Lodge equation

\begin{equation}
  \boldsymbol{\tilde{\sigma}} = -\mu \int_{0}^t e^{-(t - t') / \tau} d \mathbf{B}_e(t, t')
\end{equation}

Integration by parts gives

\begin{equation}
  \boldsymbol{\tilde{\sigma}} = -\mu (\mathbf{I} - e^{-t  / \tau}  \mathbf{B}_e(t, 0)   - \frac{1}{\tau}  \int_{0}^t \mathbf{B}_e(t, t')  e^{-(t - t') / \tau} d t' )
\end{equation}

Ignoring the term proportional to identity matrix (since it will only redefine the pressure) we rewrite it as a single integral

\begin{equation}
  \boldsymbol{\tilde{\sigma}} = \mu \frac{1}{\tau} ( \int_{0}^t  d t' e^{-(t - t') / \tau}  \frac{e^{-t  / \tau}}{1 - e^{-t  / \tau} }  \mathbf{B}_e(t, 0)   +   \mathbf{B}_e(t, t')   )
\end{equation}

Then using the definition of strain tensor

\begin{equation}
  \boldsymbol{\tilde{\sigma}} = \mu \frac{1}{\tau} ( \int_{0}^t  d t' e^{-(t - t') / \tau}  \frac{e^{-t  / \tau}}{1 - e^{-t  / \tau} }  \mathbf{F}_e(t, 0) \mathbf{F}_e(t, 0)^T   +   \mathbf{F}_e(t, t') \mathbf{F}_e(t, t')^T  )
\end{equation}

Using composition rule

\begin{equation}
  \boldsymbol{\tilde{\sigma}} = \mu \frac{1}{\tau}  \int_{0}^t  d t' e^{-(t - t') / \tau}  \frac{e^{-t  / \tau}}{1 - e^{-t  / \tau} }  \mathbf{F}_e(t, 0) \mathbf{F}_e(t, 0)^T   +   \mathbf{F}_e(t, 0) \mathbf{F}_e(0, t') \mathbf{F}_e(0, t')^T \mathbf{F}_e(t, 0)^T
\end{equation}

This allows us to write

\begin{equation}
  \mathcal{E}_t = \frac{\mu}{2} \frac{1}{\tau}  \int_{0}^t  d t' e^{-(t - t') / \tau}  \frac{e^{-t  / \tau}}{1 - e^{-t  / \tau} }  {\rm tr} ( \mathbf{F}_e(t, 0) \mathbf{F}_e(t, 0)^T   +   \mathbf{F}_e(t, t') \mathbf{F}_e(t, t')^T    )
\end{equation}

where

\begin{equation}
    \boldsymbol{\sigma}_t = \frac{\delta \mathcal{E}_t }{\delta  \mathbf{F}_t}  \mathbf{F}_t^T 
\end{equation}

It is convenient that $\mathcal{E}_t = \frac{1}{2} {\rm tr} \boldsymbol{\sigma}_t $. So unlike the standard Hookean case where energy is quadratic on deformation, and stress is linear, here both are quadratic.

\section{Unidirectionally growing beam}

\subsection{Derivation from general theory}

The unidirectionally growing beam has to bend to increase its length during the growth. If the growth is significant, the displacement is big as well. However, the curvature might be small if the beam is thin enough. At a particular moment of time $t_0$ consider a small element of a beam. Direct the $x$ axis along the axis of the beam, and the $y$ axis perpendicular to it (the capital $Y$ is reserved throughout for the transverse displacement of the centerline). If there are no bending, no rotation and no displacement, we expect 

\begin{eqnarray}
    x_0(t_1) &=& x_0(t_0) e^{g (t_1 - t_0)}, \\
    y_0(t_1) &=& y_0(t_0)
\end{eqnarray}

The growth here is unidirectional: the whole volumetric rate of Eq.~\ref{eq:continuity-incompressible} is carried by the beam axis, $\partial_x v_x = g$ and $\partial_y v_y = 0$, so the beam elongates as $e^{g t}$ at constant thickness. The determinant of the deformation gradient is then $e^{g (t_1 - t_0)}$, as it must be.

The displacement gradient tensor is then

\begin{eqnarray}
    F_{x x} = e^{g (t_1 - t_0)}
    \\
    F_{x y} = 0
    \\
    F_{y y} = 1
\end{eqnarray}

    Now we would like to take bending into account as well. We consider two moments of time, \(t_1\) and \(t_2\), and aim to determine \(\mathbf{F}(t_2,t_1)\). A small initial length element \(dx\), aligned along the beam axis, changes its length as follows:
\begin{eqnarray}
    dx_1 &=&  \big(1 + Y''(x, t_1) y(t_1)\big)\, dx_0 e^{g (t_1 - t_0)}, \\
    dx_2 &=&  \big(1 + Y''(x, t_2) y(t_2)\big) e^{g (t_2 - t_0)} dx_0,
\end{eqnarray}
where \(y(t)\) denotes the transverse coordinate of the material point at time \(t\).

Then we obtain
\begin{eqnarray}
 F_{x x} =    \frac{ dx_2 }{d x_1}  &=&  (1 +  Y''(x, t_2)  y(t_2) - Y''(x, t_1) y(t_1)) e^{g (t_2 - t_1)} .
\end{eqnarray}

Therefore the elastic part of the displacement gradient tensor is 

\begin{eqnarray}
 F^{e}_{x x} =    \frac{ dx_2 }{d x_1} e^{g (t_1 - t_2)}  &=&  (1 +  Y''(x, t_2)  y(t_2) - Y''(x, t_1) y(t_1))  .
\end{eqnarray}

The same as for only bending, non growing case. This makes total physical sense, since the growth of this element is essentially unconstrained and results in bending rather than in thickening of the beam. 

Then we have the same elastic part of the strain tensor as for non-growing case

\begin{eqnarray}
    B_{xx}^{(e)}(t_2, t_1)  &=& 1
    + 2 \big(  Y''(x, t_2) - Y''(x, t_1)\big) y(t_2),
   \\
   B_{yy}^{(e)}(t_2, t_1)  &=&1
  - 2 \big( Y''(x, t_2) -Y''(x, t_1)\big)  y(t_2).
\end{eqnarray}

Finally, we use the Lodge equation to determine the stress:
\begin{equation}
  \boldsymbol{\tilde{\sigma}} = -\mu \int_{t_0}^t e^{-(t - t') / \tau}\, d \mathbf{B}_e (t, t').
\end{equation}

To study the beam dynamics, we use the torque balance equation (see~\cite{Landau1986}):  
\begin{equation}
T'' +  ( F Y'') + K = 0,
\end{equation}

where \(T\) is the torque acting on the beam element,  
\begin{equation}
    T = \int dy\, y \sigma_{xx} =  \int dy\, y \,(\tilde{\sigma}_{xx} - p),
\end{equation}
\(F\) is the compressive load,  
\begin{eqnarray}
    F = \int dy \, (\sigma_{xx} - \sigma_{yy}),
\end{eqnarray}
and \(K\) is the external force acting on each element of the beam. Here we consider the case when there is no external force actin on the beam.

The integrals over \(y\) are taken across the cross-section of the beam, with \(y = 0\) corresponding to the midline of the beam at the current moment of time.

In these formulas, we use the usual relation 
\(\sigma_{xx} = \tilde{\sigma}_{xx} - p\).  
To determine \(p\), we impose the boundary condition at the upper and lower boundaries of the beam,  
\(\sigma_{yy} = 0\),  

This gives
\[
p = \tilde{\sigma}_{yy} .
\]

and

\begin{equation}
     T = \int dy\, y \sigma_{xx} =  \int dy\, y \,(\tilde{\sigma}_{xx} -  \tilde{\sigma}_{yy}),
\end{equation}
and 
\begin{eqnarray}
    F = \int dy \,(\tilde{\sigma}_{xx} -  \tilde{\sigma}_{yy}),
\end{eqnarray}

Explicit calculation may give a deceiving impression that $F = 0$. Actually, this is correct only up to linear order in curvature. To correctly evaluate $F$ instead we consider it as a Lagrange multiplier enforcing that the length of the beam is growing with the rate $g$ while end-to-end distance remains the same.

Using Lodge equation

\begin{equation}
  \boldsymbol{\tilde{\sigma}} = -\mu \int_{t_0}^t e^{-(t - t') / \tau}\, d \mathbf{B}(t, t').
\end{equation}

we get for the torque

\begin{equation}
  T = - I \int_{t_0}^t e^{-(t - t') / \tau}\, d (B_{xx} - B_{yy} ).
\end{equation}

where  we introduce the bending modulus, in which the elastic modulus of the material is absorbed together with the second moment of the cross-section,
\begin{eqnarray}
      I = 2 \mu \int_{-a}^{a} dy\, y^2  = \frac{4 \mu  a^3 }{3},
      \label{eq:bending-modulus}
\end{eqnarray}

 and $2 a$ is the thickness of the beam. Throughout the paper $I$ denotes this combination, so that no separate elastic modulus appears in the torque.

So 

\begin{equation}
  T = I \int_{t_0}^t e^{-(t - t') / \tau}\, d \frac{1}{R(t')} .
\end{equation}

where $R$ is the radius of the curvature, $1 / R = Y''$ in the comoving coordinates. 
Rather than writing $R$ in terms of comoving coordinate frame, it is more convenient to introduce angle $\theta$ between the beam axis and the initial direction of the beam axis. Then

\begin{eqnarray}
   \frac{1}{R} =  \frac{d \theta}{d s_{current}}
\end{eqnarray}

where $s_{current}$ is the arclength of the beam in the current configuration. In terms of the arclength of the beam at the initial moment of time we have

\begin{eqnarray}
    s_{current} = e^{g t} s_{initial}
\end{eqnarray}

so

\begin{eqnarray}
   \frac{1}{R} =  e^{- g t} \frac{d \theta}{d s_{initial}}
   \label{eq:curv-comoving}
\end{eqnarray}

Then, in the initial coordinate frame we have 

\begin{equation}
  T(s,t) = I \int_{t_0}^t e^{-(t - t') / \tau}  d e^{-g t'} \theta'(s,t') .
\end{equation}

Integrating it by parts we get (assuming an undeformed state at $t_0 = 0$)

\begin{equation}
  T(s,t) = I \left(  e^{-g t} \theta'(s,t) -  \frac{1}{\tau} \int_{t_0}^t d t' e^{-(t - t') / \tau}   e^{-g t'} \theta'(s,t') \right) .
\end{equation}

Or, rewriting it as a single integral

\begin{equation}
  T(s,t) = I    \frac{1}{\tau} \int_{t_0}^t d t' e^{-(t - t') / \tau} \left( \frac{
  1}{1 - e^{-t/\tau} } e^{-g t} \theta'(s,t) -  e^{-g t'} \theta'(s,t') \right) .
\end{equation}

or

\begin{equation}
  T(s,t) = I    \frac{1}{\tau} \int_{t_0}^t d t' e^{-(t - t') / \tau} e^{-g t} \frac{
  1}{1 - e^{-t/\tau} }  \left(
\theta'(s,t) - (1 - e^{-t/\tau} )  e^{-g (t' - t)} \theta'(s,t') \right) .
\label{eq:torque-single-int}
\end{equation}

\subsection{Effective Energy functional}

To compare with energy based approach in \cite{goldstein2006} , we  need to write

\begin{equation}
  T(s,t) = \frac{\delta \mathcal{E}[\theta, \theta', t]}{\delta \theta'(s,t)}.
\end{equation}

we assume here $t$ as a parameter, while $s$ is a variable. I.e., instead of one function $\theta(s,t)$ that depends on two variables we consider a set of functions $\theta_t(s)$ depending on only one variable $s$. Then we reproduce the correct equation for $T$, Eq.~\ref{eq:torque-single-int}, setting

\begin{equation}
\mathcal{E}[\theta, \theta', t] = I  \int_0^{L_0} ds    \frac{1}{2 \tau} \int_{t_0}^t d t' e^{-(t - t') / \tau} e^{-g t} \frac{
  1}{1 - e^{-t/\tau} }  \left(
\theta'(s,t) - (1 - e^{-t/\tau} )  e^{-g (t' - t)} \theta'(s,t') \right)^2
\end{equation}

So

\begin{eqnarray}
\mathcal{E}[\theta, \theta', t] =   \int_0^{L_0} ds   \frac{1}{\tau} \int_0^t d t'  \frac{
  e^{-(t - t') / \tau} e^{-g t}}{1 - e^{-t/\tau} }  \mathcal{H}[\theta, \theta', t, t']
\end{eqnarray}
with

\begin{eqnarray}
 \mathcal{H}[\theta, \theta', t, t'] = \frac{I}{2 } \left(
\theta'(s,t) - (1 - e^{-t/\tau} )  e^{-g (t' - t)} \theta'(s,t') \right)^2
\label{eq:H-functional}
\end{eqnarray}

It is instructive to rewrite this functional in the physical variables used in the main text: the real curvature and the current arclength. The curvature is the derivative of the angle with respect to the \emph{current} arclength, so by Eq.~\ref{eq:curv-comoving} it is related to the derivative with respect to the initial one by $\kappa(s,t) = e^{-g t} \theta'(s,t)$. Substituting $\theta'(s,t) = e^{g t} \kappa(s,t)$ and $\theta'(s,t') = e^{g t'} \kappa(s,t')$, the bracket collapses,
\begin{eqnarray}
\theta'(s,t) &-& (1 - e^{-t/\tau} )  e^{-g (t' - t)} \theta'(s,t')
\nonumber \\
&=& e^{g t} \left( \kappa(s,t) - (1 - e^{-t/\tau} ) \kappa(s,t') \right) ,
\label{eq:bracket-kappa}
\end{eqnarray}
so that squaring it produces a factor $e^{2 g t}$. At the same time the integration measure transforms as $ds_{\rm current} = e^{g t} ds$, i.e.\ $ds = e^{-g t} ds_{\rm current}$, and the integration domain becomes the current length $L(t) = L_0 e^{g t}$. Collecting the three growth factors --- $e^{-g t}$ from the prefactor, $e^{2 g t}$ from the squared bracket, and $e^{-g t}$ from the measure --- they cancel identically, $e^{-g t} \cdot e^{2 g t} \cdot e^{-g t} = 1$, and we obtain the form quoted in the main text,
\begin{equation}
\mathcal{E}_t = \frac{I}{2 \tau} \int_0^{L(t)} ds_{\rm current}    \int_{t_0}^t d t' \, \frac{ e^{-(t - t') / \tau} }{1 - e^{-t/\tau} }  \left(
\kappa(s,t) - \tilde{\kappa}(t,t') \right)^2 ,
\qquad
\tilde{\kappa}(t,t') = (1 - e^{-t/\tau} ) \, \kappa(s,t') .
\label{eq:energy-kappa}
\end{equation}
Written in the physical variables, growth has disappeared from the functional altogether: it enters only through the domain of integration $L(t)$ and through the memory of the curvature.

\subsection{Differential equation for numerical solution}
\label{sec:beam-numerics}

To get a numerical solution we start with

\begin{equation}
  T(s,t) = I \left(  e^{-g t} \theta'(s,t) -  \frac{1}{\tau} \int_{t_0}^t d t' e^{-(t - t') / \tau}   e^{-g t'} \theta'(s,t') \right) .
  \label{eq:torque-byparts}
\end{equation}

And rewrite it as

\begin{equation}
 \frac{1}{\tau} \int_{t_0}^t d t' e^{-(t - t') / \tau}   e^{-g t'} \theta'(s,t')   =    e^{-g t} \theta'(s,t) - \frac{T(s,t)}{I} .
 \label{eq:torque-int-isolated}
\end{equation}

And take a time derivative from the torque

\begin{equation}
 \frac{d}{dt} T(s,t) = I  \left(    - g e^{-g t} \theta'(s,t) + e^{-g t}  \frac{d}{dt}  \theta'(s,t)   -  \frac{1}{\tau} (    e^{-g t} \theta'(s,t) - \frac{1}{\tau}   \int_{t_0}^t d t' e^{-(t - t') / \tau}   e^{-g t'} \theta'(s,t') ) \right) .
 \label{eq:torque-dt-raw}
\end{equation}

We can represent the integral as a torque using Eq.~\ref{eq:torque-int-isolated} and get

\begin{equation}
 \frac{d}{dt} T(s,t) = I  \left(    - g e^{-g t} \theta'(s,t) + e^{-g t}  \frac{d}{dt}  \theta'(s,t)   -  \frac{1}{\tau} (    e^{-g t} \theta'(s,t) - ( e^{-g t} \theta'(s,t) - \frac{T(s,t)}{I} ) \right) .
\end{equation}

So

\begin{equation}
 \frac{d}{dt} T(s,t) =    I (  - g e^{-g t} \theta'(s,t) + e^{-g t}  \frac{d}{dt}  \theta'(s,t) )  -  \frac{1}{\tau}   T(s,t)   .
\end{equation}

Or

\begin{equation}
 \frac{d}{dt} T(s,t) =    I \frac{d}{dt}  (   e^{-g t} \theta'(s,t) )  -  \frac{1}{\tau}   T(s,t)   .
 \label{eq:torque-diff}
\end{equation}

To obtain the beam dynamics, we need to solve Eq.~\ref{eq:torque-diff} numerically together with the force balance equation at each moment of time

\begin{equation}
e^{-g t} \frac{dT}{ds} = F \sin \theta(s,t),
\label{eq:force-bal}
\end{equation}

where the factor $e^{-g t}$ comes from taking the derivative $\frac{d}{ds}$ in the initial coordinates. However, since $F$ is a time dependent Lagrange multiplier, it just rescales $F$ and we can  omit it. We therefore work from now on with the rescaled load
\begin{equation}
    \tilde{F}(t) = e^{g t} F(t) ,
    \label{eq:Ftilde}
\end{equation}
in terms of which the force balance reads simply $\frac{dT}{ds} = \tilde{F} \sin \theta$. Note that $\tilde{F}$ is not the physical load; the physical one is recovered as $F = e^{-g t} \tilde{F}$.

To solve Eqs.~\ref{eq:torque-diff} and \ref{eq:force-bal} numerically, it is convenient to introduce a new function, the deviation from pure elasticity,

\begin{eqnarray}
    q(s,t) = \frac{d}{ds} ( T(s,t) -  I e^{-g t} \theta'(s,t) )
    \label{eq:q-def}
\end{eqnarray}

For the purely elastic case $q = 0$ always. For viscoelastic, we rewrite equations in the form

\begin{eqnarray}
    \frac{d q}{d t} = - \frac{\tilde{F}}{\tau} \sin \theta
    \label{eq:q-evol}
    \\
    I e^{- g t} \frac{d^2 \theta}{d s^2} - \tilde{F} \sin \theta  + q = 0
    \label{eq:q-balance}
\end{eqnarray}

Eq.~\ref{eq:q-evol} follows by differentiating the torque equation Eq.~\ref{eq:torque-diff} with respect to $s$ and inserting it into the definition Eq.~\ref{eq:q-def}: the terms $I \frac{d}{d t} ( e^{-g t} \theta'' )$ cancel identically, leaving $\frac{d q}{d t} = - \frac{1}{\tau} \frac{d T}{d s} = - \frac{\tilde{F}}{\tau} \sin \theta$. Eq.~\ref{eq:q-balance} follows from Eq.~\ref{eq:q-def} together with the force balance Eq.~\ref{eq:force-bal}, which give $q = \tilde{F} \sin \theta - I e^{-g t} \theta''$. Both relations are exact and do not rely on a small-angle expansion.

To make the dynamics more stable from a numerical point of view, we replace the condition of pinned ends by an elastic spring with a large spring constant, so that the distance between the ends of the beam is almost constant. $\tilde{F}$ then has an expression in terms of the end to-end distance $\tilde{F} = k(L_0 - e^{g t} \int_0^{L_0} ds \cos \theta )$.

Eq.~\ref{eq:q-balance} does not contain a time derivative --- it is just a boundary value problem at each moment of time. To solve the system numerically, we find $q$ at the next moment of time using Eq.~\ref{eq:q-evol}, and use this $q$ to determine $\theta$. We look for the solution for $\theta$ close to the solution at the previous moment of time. However, when resonance occurs
\begin{eqnarray}
  -\frac{\tilde{F} \cos \theta}{   I e^{- g t} }=  \left( \frac{\pi n}{L_0} \right)^2
  \label{eq:resonance}
\end{eqnarray}

there is no solution close to the previous one. To find the next state in this situation, we introduce fictitious time $\tilde{t}$ and evolve

\begin{eqnarray}
 \zeta \frac{d \theta}{d \tilde{t} } =  I e^{- g t} \frac{d^2 \theta}{d s^2} - \tilde{F} \sin \theta  + q
\end{eqnarray}

where $\zeta$ is a fictitious drag coefficient that sets only the rate of the relaxation, not the state it relaxes to, until we find a new stationary state, which is far from the previous one. That is how snaps happen.

\subsection{Small angle limit: analytical solution}
\label{sec:small-angle}

Immediately after the growth starts the beam is still almost straight, and the equations of motion can be linearized. We show here that in this regime the shape of the beam is fixed entirely by its current length: the parameter $g \tau$ drops out of the shape and survives only in the compressive load.

We start from the differential form of the torque equation, Eq.~\ref{eq:torque-diff}, together with the force balance at each moment of time, Eq.~\ref{eq:force-bal},
\begin{eqnarray}
    \frac{d T}{d t} &=& I \frac{d}{d t} \left( e^{-g t} \theta'(s,t) \right) - \frac{1}{\tau} T(s,t) ,
    \\
    e^{-g t} \frac{d T}{d s} &=& F \sin \theta(s,t) ,
\end{eqnarray}
the latter of which, in terms of the rescaled load $\tilde{F} = e^{g t} F$ (Eq.~\ref{eq:Ftilde}), reads simply $d T / d s = \tilde{F} \sin \theta$. Differentiating the torque equation with respect to $s$, using the force balance to eliminate $d T / d s$ in favour of the load, and linearizing for small angles, $\sin \theta \approx \theta$, we obtain a single linear equation for the angle,
\begin{equation}
    \frac{d}{d t} \left( \tilde{F} \theta \right) = I \frac{d}{d t} \left( e^{-g t} \theta''(s,t) \right) - \frac{1}{\tau} \tilde{F} \theta .
    \label{eq:sa-lin}
\end{equation}

The ends of the beam are pinned, so no torque is applied to them, $T(0,t) = T(L_0, t) = 0$ at all times. Since the beam is undeformed at $t_0 = 0$, the integral expression for the torque then requires
\begin{equation}
    \theta'(0,t) = \theta'(L_0,t) = 0 .
\end{equation}
This is satisfied by
\begin{equation}
    \theta(s,t) = A(t) \cos \frac{\pi n s}{L_0} ,
    \label{eq:sa-ansatz}
\end{equation}
for which $\theta'' = - (\pi n / L_0)^2 \theta$, so that Eq.~\ref{eq:sa-lin} is satisfied identically in $s$ and collapses to a single ordinary differential equation relating the amplitude $A(t)$ and the load $\tilde{F}(t)$. In terms of $G(t) = \tilde{F}(t) A(t)$ it reads
\begin{equation}
    \dot{G} + \frac{G}{\tau} = - I \left( \frac{\pi n}{L_0} \right)^2 \frac{d}{d t} \left( e^{-g t} A \right) .
    \label{eq:sa-G}
\end{equation}

The amplitude itself is fixed not by the dynamics but by the constraint that the end-to-end distance stays $L_0$ while the arclength grows as $L(t) = L_0 e^{g t}$:
\begin{equation}
    e^{g t} \int_0^{L_0} ds \, \cos \theta(s,t)  = L_0 .
\end{equation}
Expanding to quadratic order in $\theta$, we get
\begin{equation}
   \int_0^{L_0} ds \, \theta^2(s,t)  = 2 L_0 \left( 1 - e^{-g t} \right) ,
\end{equation}
and substituting the ansatz Eq.~\ref{eq:sa-ansatz}, for which $\int_0^{L_0} \theta^2 ds = A^2 L_0 / 2$ for any $n$, we obtain
\begin{equation}
    A(t) = 2 \sqrt{1 - e^{-g t}} = 2 \sqrt{1 - \frac{L_0}{L(t)}} .
    \label{eq:sa-amplitude}
\end{equation}

This is the central result of this subsection. The amplitude depends only on the ratio of the current length to the initial one, and is independent of $g \tau$, of the bending modulus $I$, and of the load $F$. In terms of the transverse displacement of the center line, $Y' = \tan \theta \approx \theta$, the shape is a sine wave
\begin{equation}
    Y(s,t) = \frac{A(t) L_0}{\pi n} \sin \frac{\pi n s}{L_0} .
\end{equation}

The mode number $n$ is selected by the load: the right hand side of Eq.~\ref{eq:sa-G} is proportional to $n^2$, hence so is $G$, and hence so is the load $F = e^{-g t} G / A$. Thus $F_n = n^2 F_1$ — every higher mode requires a larger compressive load, so the beam buckles into the $n = 1$ mode. From now on we set $n = 1$.

With $A(t)$ now known, Eq.~\ref{eq:sa-G} determines the load. It is solved by
\begin{equation}
    G(t) = - I \left( \frac{\pi }{L_0} \right)^2 \int_0^t dt' \, e^{-(t - t')/\tau} \frac{d}{dt'} \left( e^{-g t'} A(t') \right) .
\end{equation}
Using $e^{-g t} A = 2 e^{-g t} \sqrt{1 - e^{-g t}}$ and
\begin{equation}
    \frac{d}{dt} \left( e^{-g t} A \right) = - g \frac{e^{-g t} \left( 2 - 3 e^{-g t} \right)}{\sqrt{1 - e^{-g t}}} ,
\end{equation}
and recovering the rescaled load as $\tilde{F} = G / A$, and the physical one as $F = e^{-g t} \tilde{F} = e^{-g t} G / A$, we obtain
\begin{equation}
    F(t) = I \left( \frac{\pi }{L_0} \right)^2 \frac{g \, e^{-g t}}{2 \sqrt{1 - e^{-g t}}} \int_0^t dt' \, e^{-(t - t')/\tau} \frac{e^{-g t'} \left( 2 - 3 e^{-g t'} \right)}{\sqrt{1 - e^{-g t'}}} .
    \label{eq:sa-force-full}
\end{equation}
Unlike the shape, the load does depend on $g \tau$.

It is instructive to check the two limits. In the elastic limit $g \tau \gg 1$ the exponential kernel is unity over the whole integration range, the integral telescopes to $e^{-g t} A(t) - A(0) = e^{-g t} A(t)$, and
\begin{equation}
    F = - I \left( \frac{\pi }{L_0} \right)^2 e^{- 2 g t} = - \frac{ I \pi^2}{L^2(t)} ,
\end{equation}
which is exactly the Euler buckling load for a beam of the current length $L(t)$: the growing elastic beam stays right at the buckling threshold set by its instantaneous length. In the opposite, viscous limit $g \tau \ll 1$, the kernel is sharply peaked and $G \approx - \tau I (\pi/L_0)^2 \frac{d}{dt} ( e^{-g t} A )$, giving
\begin{equation}
    F =  I \left( \frac{\pi }{L_0} \right)^2  \frac{ g \tau \, e^{- 2 g t} \left( 2 - 3 e^{-g t} \right) }{2 \left( 1 - e^{-g t} \right)} ,
\end{equation}
valid for $\tau \ll t$, where the quasi-static approximation applies.

Finally, we note the range of validity. The linearization requires $A \ll 1$, i.e.\ $g t \ll 1$, where $A \approx 2 \sqrt{g t}$ — the familiar square-root growth of the buckling amplitude with the excess length. In this window both limits above reduce to
\begin{eqnarray}
    F_{\rm elastic} &\approx& - I \left( \frac{\pi }{L_0} \right)^2 ,
    \\
    F_{\rm viscous} &\approx& - I \left( \frac{\pi }{L_0} \right)^2 \frac{\tau}{2 t} ,
\end{eqnarray}
so the two coincide at $t \sim \tau$, as expected. Once $g t \sim 1$ the amplitude becomes of order unity, the quadratic expansion of the constraint and the linearization of $\sin \theta$ both fail, the sine wave is no longer a solution, and $g \tau$ enters the shape of the beam itself — which is the regime reached by the numerical solution of Sec.~\ref{sec:beam-numerics}.

\subsection{Beam with growing thickness}

If we would like to incorporate the growth in thickness for the beam (so it is growing with two different growth rates in longitudinal and transverse directions, but still the stress is weak and the thickening due to compression is negligible), the derivation remains almost the same. We still have the same elastic part of the strain

\begin{eqnarray}
    B_{xx}^{(e)}(t_2, t_1)  &=& 1
    + 2 \big(  Y''(x, t_2) - Y''(x, t_1)\big) y(t_2),
   \\
   B_{yy}^{(e)}(t_2, t_1)  &=&1
  - 2 \big( Y''(x, t_2) -Y''(x, t_1)\big)  y(t_2).
\end{eqnarray}

since there is no thickening due to elastic deformation, only bending. The only thing that gets modified is torque

where \(T\) is the torque acting on the beam element,  
\begin{equation}
    T = \int dy\, y \sigma_{xx} =  \int dy\, y \,(\tilde{\sigma}_{xx} - p),
\end{equation}

since the integration limits are now changing due to the growth of the beam, so it leads to increasing moment of inertia $I$ with time. In the previous derivation we freely commuted $I $ with the time derivative, which we can not do here. Doing it properly, we obtain

\begin{equation}
 \frac{d}{dt} \left( \frac{T(s,t)}{I(t)} \right) =     \frac{d}{dt}  (   e^{-g t} \theta'(s,t) )  -  \frac{1}{\tau}   \left( \frac{T(s,t)}{I(t)} \right).
\end{equation}

and the force balance remains the same

\begin{equation}
e^{-g t} \frac{dT}{ds} = F \sin \theta(s,t),
\end{equation}

Rescaling $\hat{T} = T/ I$, and $\hat{F} = F / I $ (not to be confused with the rescaled load $\tilde{F} = e^{g t} F$ of Eq.~\ref{eq:Ftilde}), we obtain the same equations as Eqs.~\ref{eq:torque-diff} and \ref{eq:force-bal} earlier

\begin{equation}
 \frac{d}{dt} \hat{T}  =     \frac{d}{dt}  (   e^{-g t} \theta'(s,t) )  -  \frac{1}{\tau}    \hat{T}.
\end{equation}

and

\begin{equation}
e^{-g t} \frac{d \hat{T} }{ds} = \hat{F} \sin \theta(s,t),
\end{equation}

The equations now do not depend on $I $. This means that for any $I$, including a growing one, the shape of the beam will be the same, as long as the slender beam assumption holds (the curvature radius is much bigger than the thickness).

\section{Wrinkling of a growing layer}

Previously, we considered the case where the deformation of each element of the slender beam is due to the growth, and the deformation due to the stress is negligible. Here we consider another limit, in which the beam only starts to bend but already gets thicker, and explore the growth of the instability at these initial moments. We consider a beam of uniformly growing material clamped between two walls.

This  geometry serves as a minimal model for the rim of a biofilm: the outer layer is growing while remaining attached to a non-growing inner core. In such a scenario, the radius of the rim cannot expand freely, as it remains connected to the bulk, but the circumference must increase due to growth. This mismatch naturally leads to buckling instabilities, manifested as radial wrinkles.

In the limit of large radius, where the curvature of the rim is small compared to the other relevant length scales, the problem reduces to the classical mechanics setup of a growing beam confined between two rigid boundaries. We therefore begin by analyzing this beam geometry, which will allow us to highlight the role of viscoelasticity in growth-induced wrinkling.

We assume that the rigid walls block motion in the $x$-direction but do not exert any tangential friction on the beam (i.e.\ $\xi = 0$ in the terminology of the previous section). The beam is taken to be undeformed at $t=0$, and growth begins at that instant. We look for solutions in the small-wrinkling (small-deflection) regime, so that the beam remains nearly straight and curvature effects enter only as higher-order corrections.

To leading order (the straight-beam approximation, i.e.\ zero order in curvature), the displacement is purely transverse: only the $y$-coordinate changes in time while the $x$-coordinate of each material point remains fixed. Denoting the reference (material) coordinate of a point by $x_0(0)$ and its current coordinates by $x_0(t), y_0(t)$, we obtain

\begin{eqnarray}
    x_0(t) &=& x_0(0), \\
    y_0(t) &=& y_0(0)\, e^{g t},
\end{eqnarray}

Here the partitioning of the growth is the opposite one: the walls block elongation along $x$, so the whole volumetric rate of Eq.~\ref{eq:continuity-incompressible} is taken up by the transverse direction, $\partial_y v_y = g$ and $\partial_x v_x = 0$. The layer therefore thickens as $e^{g t}$ at fixed length, and again $\det \mathbf{F} = e^{g t}$.

\begin{figure}
    \centering
    \includegraphics[width=0.8\linewidth]{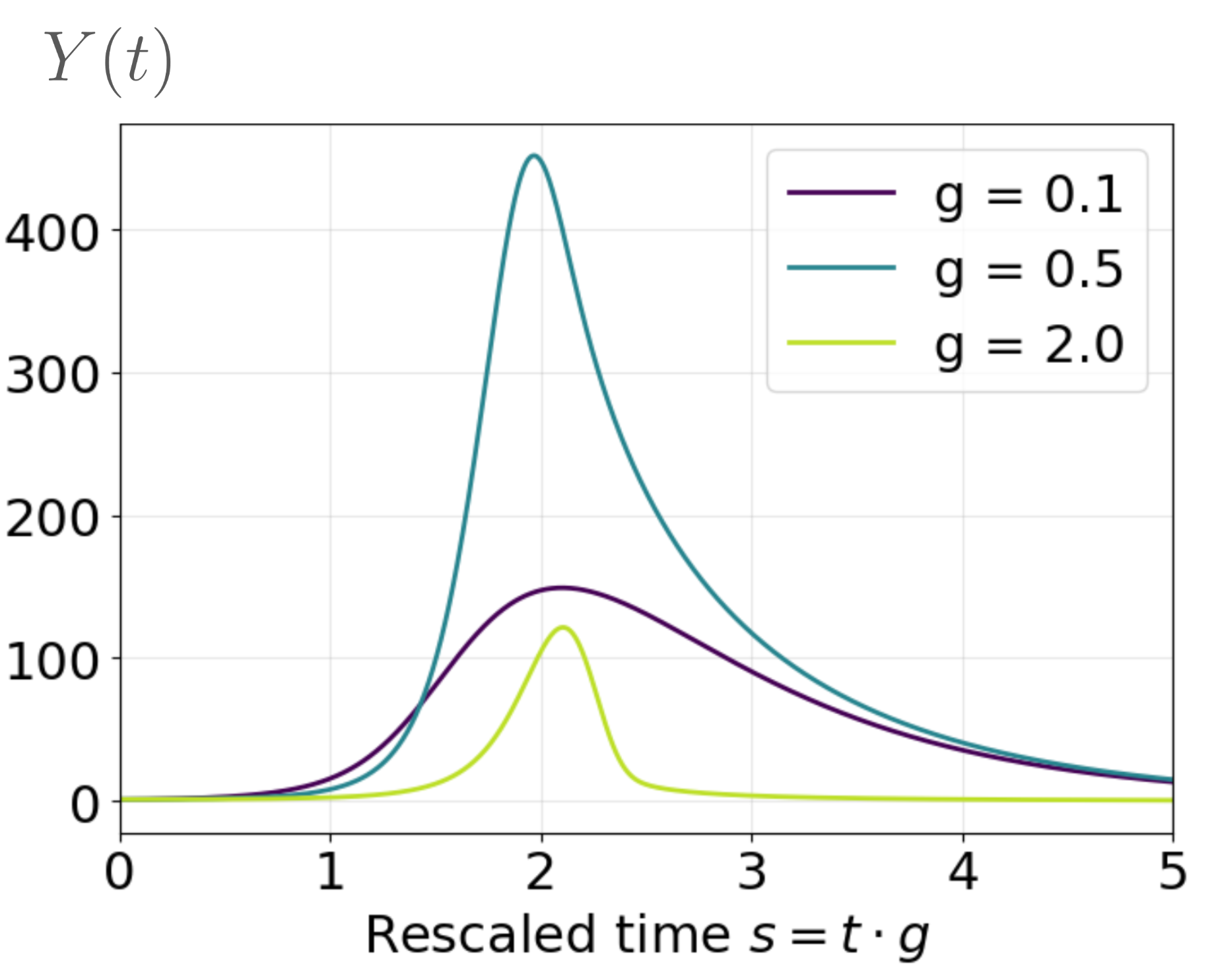}
     \includegraphics[width=0.8\linewidth]{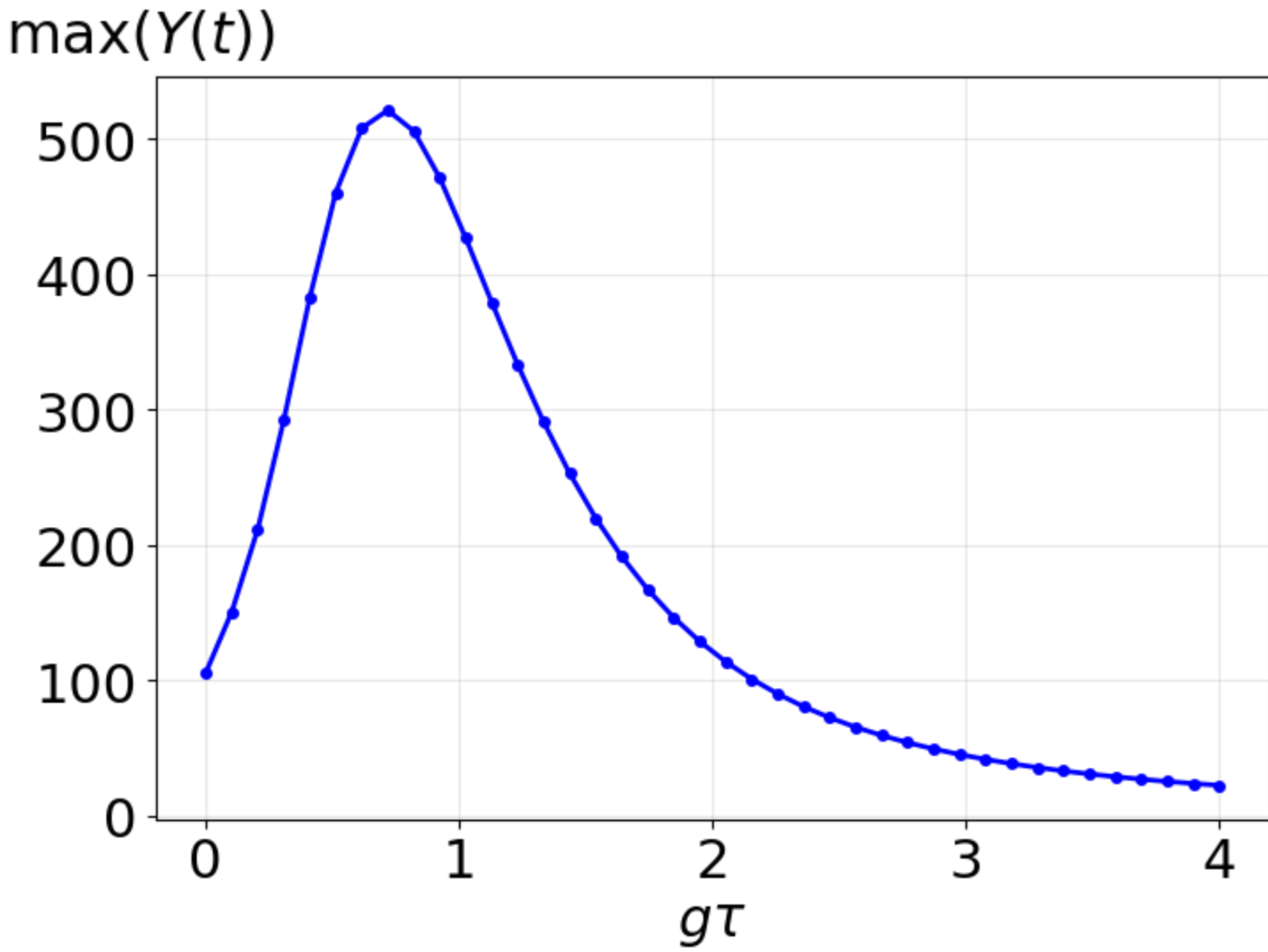}
    \caption{Wrinkling of a growing layer confined between two frictionless walls. Top: the amplitude $Y_k(t)$ of the wrinkling mode $Y(x,t) = Y_k(t) \sin (k x)$ (Eq.~\ref{eq:wrinkle-ansatz}), measured in units of its initial value $Y_k(0)$, as a function of the rescaled time $g t$, for three growth rates $g$ (in units of $1 / \tau$). Because the analysis is linear in the perturbation, the amplitude is only defined up to this initial value. For each $g$ the amplitude reaches a maximum and then decreases, as the growing thickness of the layer makes it progressively harder to bend. Bottom: the maximal amplitude $\max_t Y_k(t)$, in the same units, as a function of $g \tau$. Intermediate values, $g \tau \sim 1$, produce the largest wrinkles.}
    \label{fig:curves}
\end{figure}

To calculate the first-order correction and explore how quickly the instability develops in the wrinkling beam, we follow the classical beam-buckling analysis in \cite{Landau1986} and generalize it to the viscoelastic growing case. We assume that growth starts at the moment of time \(t = 0\), with the beam relaxed before that. Let \(Y(x,t)\) denote the transverse coordinate of the central line of the undeformed beam after deformation. We assume that the dependence of \(Y\) on \(x\) is slow compared to the other parameters of the problem. 

We consider two moments of time, \(t_1\) and \(t_2\), and aim to determine \(\mathbf{F}(t_2,t_1)\). A small initial length element \(dx\), aligned along the beam axis, changes its length as follows:
\begin{eqnarray}
    dx_1 &=&  \big(1 + Y''(x, t_1)\, y(t_1)\big)\, dx_0, \\
    dx_2 &=&  \big(1 + Y''(x, t_2)\, y(t_2)\big)\, dx_0,
\end{eqnarray}
where \(y(t)\) denotes the transverse coordinate of the material point at time \(t\).

In the approximation of small curvature (small \(Y''\)), the dominant contribution to \(y\) comes from growth, so that 
\[
y(t) = e^{g t} y(0).
\]
Then we obtain
\begin{eqnarray}
    \frac{ dx_2 - dx_1}{d x_1}  &=&  \big( Y''(x, t_2)\, e^{g (t_2 - t_1)} - Y''(x, t_1) \big)\, y(t_1).
\end{eqnarray}

Using the incompressibility condition \(\det F = e^{g t}\), and the fact that at the boundary \(\sigma_{xy} = 0\) (which is satisfied by setting \(B_{xy} = 0\) due to the Lodge equation), we obtain the deformation

\begin{eqnarray}
  x(t_2) &=& x(t_1) +   \left(  e^{g (t_2 - t_1)} Y''(x, t_2) - Y''(x, t_1) \right) y(t_1)\, x(t_1),
  \\
  y(t_2)  &=&  e^{g (t_2 - t_1)} \left( y(t_1) -  \left( e^{g (t_2 - t_1)} Y''(x, t_2) - Y''(x, t_1) \right) \frac{y^2(t_1) + x^2(t_1)}{2} \right).
\end{eqnarray}

We now use the deformation to calculate the components \(B_{xx}^{(e)}\) and \(B_{yy}^{(e)}\) of the elastic part of the strain tensor,

\begin{eqnarray}
    B_{xx}^{(e)}(t_2, t_1)  &=& e^{- g (t_2 - t_1)}
    + 2 \big( e^{g t_2} Y''(x, t_2) -e^{g t_1} Y''(x, t_1)\big)\, e^{-g t_1} y(t_2)\, e^{- 2 g (t_2 - t_1)},
   \\
   B_{yy}^{(e)}(t_2, t_1)  &=& e^{ g (t_2 - t_1)}
  - 2 \big( e^{g t_2} Y''(x, t_2) -e^{g t_1} Y''(x, t_1)\big)\, e^{-g t_1} y(t_2).
\end{eqnarray}

The component \(B_{xy}^{(e)}\) is zero by construction.

Finally, we use the Lodge equation to determine the stress:
\begin{equation}
  \boldsymbol{\tilde{\sigma}} = -\mu \int_{t_0}^t e^{-(t - t') / \tau}\, d \mathbf{B}(t, t').
\end{equation}

To study the beam dynamics, we use the torque balance equation (see~\cite{Landau1986}):
\begin{equation}
T'' +  ( F Y'') + K = 0,
\end{equation}
where \(T\) is the torque acting on the beam element,
\begin{equation}
    T = \int dy\, y \sigma_{xx} =  \int dy\, y \,(\tilde{\sigma}_{xx} - p),
\end{equation}
\(F\) is the compressive load,
\begin{eqnarray}
    F = \int dy \, (\sigma_{xx} - \sigma_{yy}),
\end{eqnarray}
and \(K\) is the external force acting on each element of the beam.  
In the overdamped regime, \(K\) is proportional to the velocity of the beam element in the direction perpendicular to the beam axis, due to friction with the surrounding medium:  
\begin{equation}
   K =  2 \chi \dot{Y}.
\end{equation}

The integrals over \(y\) are taken across the cross-section of the beam, with \(y = 0\) corresponding to the midline of the beam at the current moment of time. The beam width evolves in time as  
\[
2 a(t) = 2 a(0)\, e^{g t}.
\]

In these formulas, we use the usual relation 
\(\sigma_{xx} = \tilde{\sigma}_{xx} - p\).  
To determine \(p\), we impose the boundary condition at the upper and lower boundaries of the beam,  
\(\sigma_{yy} = \chi \dot{a}\),  
which states that the normal stress is balanced by the friction with the surrounding medium.  
This gives
\[
p = \tilde{\sigma}_{yy} - \chi \dot{a}.
\]
Substituting this into the expressions above, we obtain the compressive load \(F\):
\begin{eqnarray}
    F  = 2 a(t)  \mu g \tau \left( \frac{1}{g \tau + 1}\left(1  - e^{- t (g + 1 / \tau)}\right)
    + \frac{1}{1 - g \tau}\left(1  - e^{- t (-g + 1 / \tau)}\right) \right).
\end{eqnarray}

Similarly, for the torque \(T\) we find
\begin{eqnarray}
    T = (\omega_{xx} - \omega_{yy}) I,
\end{eqnarray}
where we introduce the bending modulus, defined as in Eq.~\ref{eq:bending-modulus} but with the time-dependent thickness,
\begin{eqnarray}
      I = 2 \mu \int dy\, y^2  = \frac{4 \mu  a(t)^3 }{3},
      \label{eq:bending-modulus-growing}
\end{eqnarray}
and the  tensor
\[
\boldsymbol{\omega} = \frac{1}{I} \int dy\, y\, \boldsymbol{\tilde{\sigma}}.
\]
Its components are given by
\begin{eqnarray}
 \omega_{xx}(t) &=& 
 - 2 e^{-t ( 2g + 1/\tau) }  \int_{t_0}^t   e^{ - t' / \tau} \frac{d}{dt'} \left( \left(  e^{g t_2} Y''(x, t_2) -e^{g t_1} Y''(x, t_1) \right)   e^{g t'} 
  \right), \\
 \omega_{yy}(t) &=& 
  2   e^{-t/\tau} \int_{t_0}^t  e^{ - t' / \tau}  \frac{d}{dt'} \left( \left(  e^{g t_2} Y''(x, t_2) -e^{g t_1} Y''(x, t_1) \right)   e^{-g t'}
   \right).
\end{eqnarray}

We can transform the integral equations for the components of tensor $\omega$ above into differential form by taking the time derivative, yielding

\begin{eqnarray}
 \dot{\omega}_{xx}(t) &=& 
  - ( 2g + 1/\tau) \omega_{xx}(t)
  +    \frac{2}{1  + g \tau}  (\dot{Y}'' + g Y'') \left( 1 +  g \tau e^{-t ( g + 1/\tau) } \right),       
   \\
 \dot{\omega}_{yy}(t) &=& 
  - \frac{1}{\tau} \omega_{yy}(t)
  -  \frac{2}{1 - g \tau} (\dot{Y}'' + g Y'')   \left(1 -  g \tau e^{-t/\tau + g t} \right).
\end{eqnarray}
We now fix the $x$-dependence of the perturbation from the boundary conditions at the walls. A frictionless wall transmits a purely normal (axial) force to the beam end, so it exerts neither a couple nor a transverse force there. The vanishing couple means that the bending moment vanishes at both walls, $T(0,t) = T(L,t) = 0$; since $T$ is a linear functional of the history of $Y''$ and the beam is undeformed at $t = 0$, this is enforced by $Y''(0,t) = Y''(L,t) = 0$. The absence of a transverse reaction excludes, in addition, a contribution linear in $x$. The admissible modes are therefore standing sine waves rather than complex exponentials,

\begin{equation}
    Y(x,t) = Y_k(t) \sin (k x), \qquad k = \frac{\pi n}{L}, \quad n = 1, 2, \ldots
    \label{eq:wrinkle-ansatz}
\end{equation}

where $L$ is the separation between the walls and $n$ counts the half-wavelengths. The remaining solution of the boundary-value problem, a uniform translation of the whole beam in the $y$-direction, is a rigid-body mode: it is opposed by no restoring force and does not wrinkle the beam, so we discard it. Writing likewise $\omega_{xx}(x,t) = m_{xx}(t) \sin (k x)$ and $\omega_{yy}(x,t) = m_{yy}(t) \sin (k x)$, the sine factors out of every equation, since the $x$-dependence enters only through $Y'' = -k^2 Y$, and we obtain a closed system of ordinary differential equations for the amplitudes:

\begin{eqnarray}
 \dot{m}_{xx}(t) &=&
  - ( 2g + 1/\tau) m_{xx}(t)
  -   k^2 \frac{2}{1  + g \tau}  (\dot{Y}_k + g Y_k) \left( 1 +  g \tau e^{-t ( g + 1/\tau) } \right),
   \\
 \dot{m}_{yy}(t) &=&
  - \frac{1}{\tau} m_{yy}(t)
  +  k^2 \frac{2}{1 - g \tau} (\dot{Y}_k + g Y_k)   \left(1 -  g \tau e^{-t/\tau + g t} \right),
  \\
 2 \chi \dot{Y}_k &=& I k^2 (m_{xx} - m_{yy}) +  k^2 ( F Y_k).
\end{eqnarray}

Solving this system numerically for various values of \(g\) using Wolfram Mathematica, we observe the behavior shown in Fig.~\ref{fig:curves}, where \(Y_k(t)\) initially increases, reaches a maximum, and then decreases.
The decrease occurs because, as the beam becomes too thick due to growth, it can no longer bend effectively.  

The maximal wrinkling amplitude depends on the growth rate in a non-monotonic way (see Fig.~\ref{fig:curves}).  
For very slow growth, stress relaxes faster than it builds up, suppressing wrinkling.  
Faster growth initially leads to a quicker increase in the wrinkling mode; however, the beam also thickens more rapidly.  
Consequently, the point at which wrinkles stop growing is reached sooner, potentially resulting in a smaller total wrinkling amplitude.  
The value of \(g \tau\) that produces maximal wrinkling is of order 1.  
This demonstrates that an intermediate growing material—neither purely elastic nor purely viscous—exhibits properties distinct from either limit, and that the dimensionless parameter controlling this behavior is \(g \tau\).

\section{Viscoelastic Growing Matter in a Channel}

One of the simplest geometrical constraints on the growth of viscoelastic matter is confinement within a channel. In natural systems, this situation arises, for example, in porous media, where cells proliferate not in free space but through the pores of a non-growing surrounding material. Here, we show that accounting for the viscoelastic properties of the growing material leads to qualitatively new effects compared to the purely viscous case.

To set the stage, we first recall the behavior of a standard viscous fluid in a channel and then extend the analysis to a growing viscoelastic fluid.

\begin{figure}
    \centering
    \includegraphics[width=0.5\linewidth]{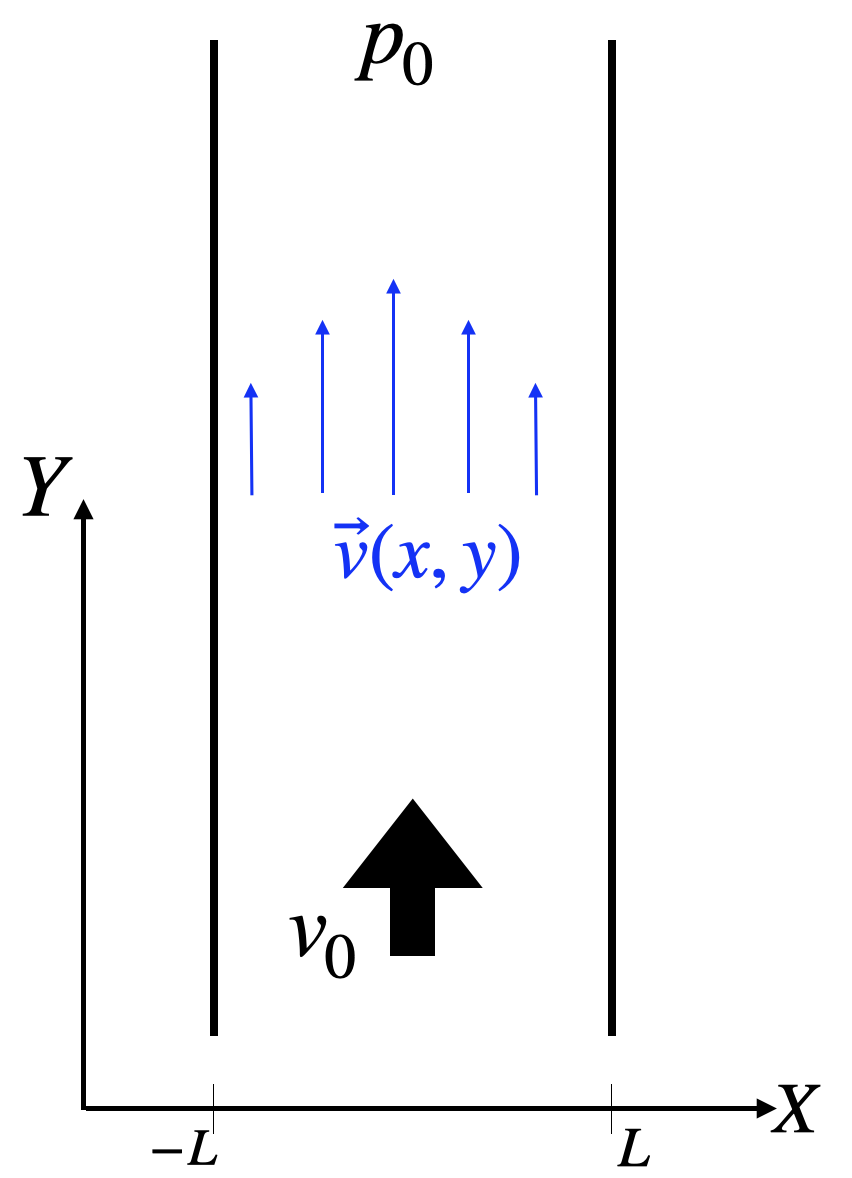}
    \caption{Illustration of the flow of the viscous fluid through the channel in two dimensions. Because of the friction to the wall, velocity at the wall is smaller than velocity at the center of the channel.}
    \label{fig:viscous-channel}
\end{figure}

We consider a 2D flow of a viscous, non-growing fluid in a channel aligned with the $y$-axis, of total width $2L$ (see Fig.~\ref{fig:viscous-channel}).

At the channel walls, we impose Navier slip boundary conditions:

\begin{equation}
\sigma_{xy}(\pm L) \pm \xi v_y(\pm L) = 0 ,
\end{equation}

where the shear stress at the boundary is proportional to the tangential velocity, with proportionality constant $\xi$. In the limit $\xi \to \infty$, the tangential velocity vanishes, corresponding to cells unable to move along the boundary. Conversely, $\xi \to 0$ represents free slip, allowing cells to slide without friction. Intermediate values of $\xi$ describe finite, nonzero friction at the walls.

\subsection{Viscous fluid in porous media. Darcy's law}

Consider a 2D flow of a viscous, non-growing fluid in a channel aligned with the $y$-axis and of width $2L$ (see Fig.~\ref{fig:viscous-channel}). Assuming force balance, $\boldsymbol{\nabla} \cdot \boldsymbol{\sigma} = 0$, the Stokes equations for an incompressible fluid in component form are

\begin{eqnarray}
\eta (\partial_x^2 + \partial_y^2) v_x &=& \partial_x p 
\\
\eta (\partial_x^2 + \partial_y^2) v_y &=& \partial_y p ,
\end{eqnarray}

supplemented by the incompressibility condition

\begin{equation}
\partial_x v_x + \partial_y v_y = 0 .
\end{equation}

At the channel walls, we impose Navier slip boundary conditions,

\begin{equation}
\sigma_{xy}(\pm L) \pm \xi  v_y(\pm L) = 0 ,
\end{equation}

where the shear stress at the boundary is proportional to the tangential velocity, with proportionality constant $\xi$. In the limit $\xi \to \infty$, the tangential velocity vanishes, corresponding to cells that cannot move along the boundary. Conversely, $\xi \to 0$ represents free slip, where cells can slide without friction along the boundary. Intermediate values of $\xi$ describe finite, nonzero friction at the walls.

Solving the Stokes equations with the incompressibility condition and Navier slip boundary conditions, we find

\begin{eqnarray}
v_x = 0 
\\
v_y =  \frac{C}{\eta} \left( \frac{x^2}{2} - \frac{\eta}{\xi} L   -   \frac{ L ^2}{ 2} \right)
    \\
     p = C y + p_0
\end{eqnarray}

where $C$ is an integration constant and $p_0$ is a reference pressure.

Averaging the velocity over the channel cross-section yields

\begin{eqnarray}    
\langle \mathbf{v} \rangle = - \left( \frac{ L}{\xi}  +   \frac{  L^2}{ 3 \eta} \right) \boldsymbol{\nabla} p
\end{eqnarray}

which is precisely Darcy’s law for flow in porous media, $\langle \mathbf{v} \rangle = - k \boldsymbol{\nabla} p$ with permeability $k = \left( \frac{ L}{\xi}  +   \frac{  L^2}{ 3 \eta} \right) $. Although $p$ is a constraint multiplier rather than the mechanical pressure, the two differ under uniform growth only by the spatially uniform offset found above, which drops out of $\boldsymbol{\nabla} p$; the pressure gradient appearing in Darcy's law is therefore the physically measured one.

\subsection{Viscous Growing Fluid in Porous Media}

In this subsection, we examine how the previous results are modified when the viscous fluid is also growing. The continuity equation now becomes

\begin{equation}
\boldsymbol{\nabla} \cdot \mathbf{v} = g,
\end{equation}

where $g$ is the uniform growth rate.

In the viscous limit for a 2D system, the constitutive equation for a growing viscoelastic fluid (Eq.~\ref{eq:const-viscoel-grow}) simplifies to

\begin{equation}
\sigma_{ij} = \eta  (\partial_i v_j + \partial_j v_i) - (p + g \eta)  \delta_{ij},
\end{equation}

In this case, the pressure term is modified compared to the standard Stokes equation. One can redefine the pressure as $p \to p + g \eta$ which restores the original form of the Stokes equation. This redefinition does not affect the physical results, since $p$ acts as a Lagrange multiplier enforcing the incompressibility (or, in this case, growth-modified continuity) constraint. 

In the case of a homogeneous growth rate, the above equations can be solved to give

\begin{eqnarray}
 p =     \eta \left( \frac{3 g  x^2}{ 4 L^2}  -  \frac{g y^2  +  2 v_0 y}{  2  L^2 / 3 +  2 \frac{\eta}{\xi}   L  } \right) + p_0   
\\
  v_x =       g  x \frac{ x^2  - L^2 - 2 \frac{\eta}{\xi}   L }{ 2 L^2}    
  \\
 v_y =  (g y  + v_0) \frac{ L^2 + 2 \frac{\eta}{\xi}   L - x^2 }{2 (   L^2 / 3 +  \frac{\eta}{\xi}   L ) }    
\end{eqnarray}

where $v_0$ is the average velocity at $y = 0$.

Averaging across the channel cross-section, this solution reproduces  Darcy’s law with the same permeability as for the non-growing fluid. In the next section, we will show that for a viscoelastic growing fluid, the relationship between flow and pressure gradient (i.e., Darcy’s law) is modified, leading to qualitatively new behavior.

\subsection{Viscoelastic Growing Fluid in Porous Media}

\begin{figure}
    \centering
    \includegraphics[width=0.8\linewidth]{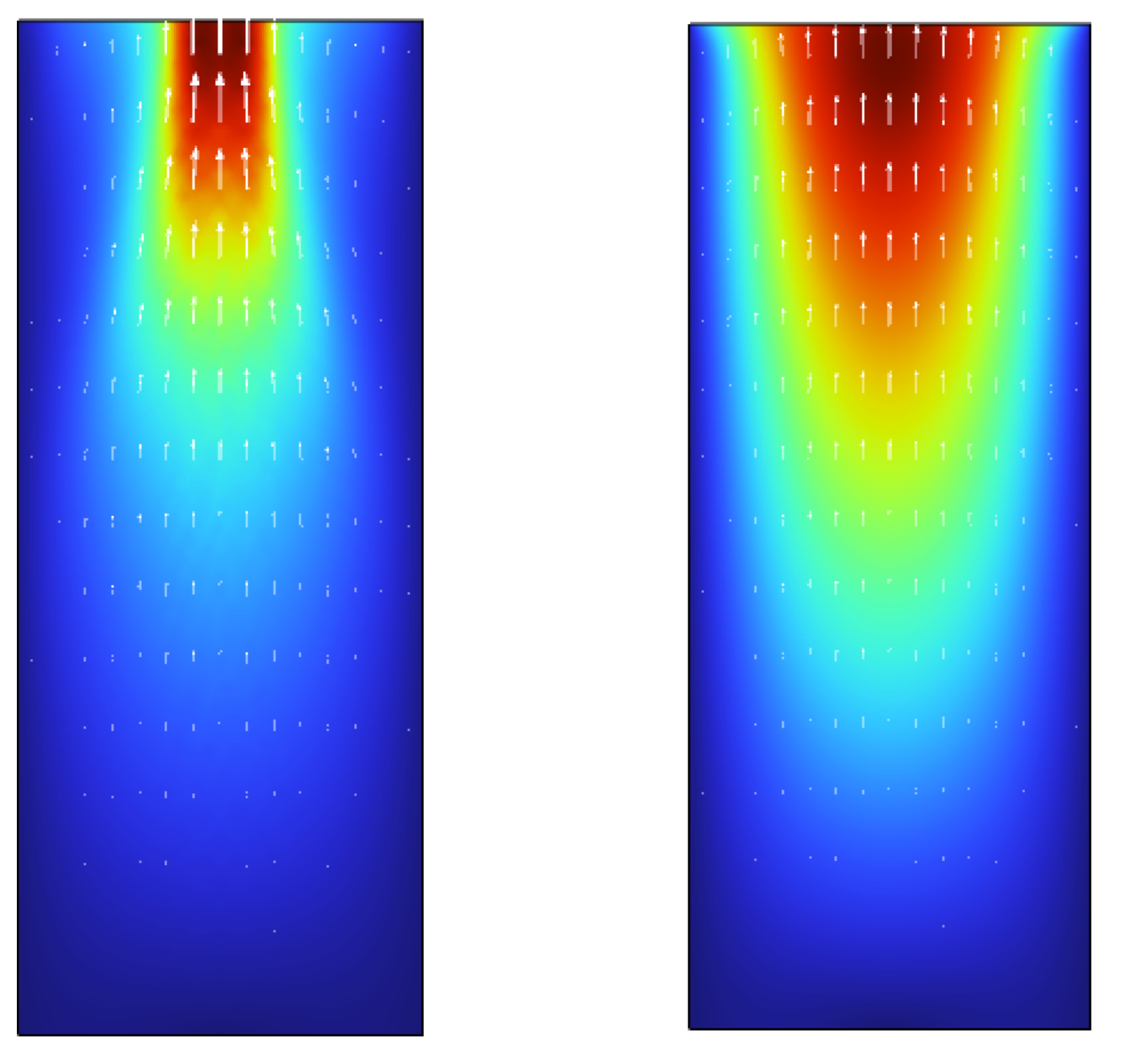}
    \caption{Numerical solution for viscoelastic growing fluid velocity profile. Arrow indicates velocity direction, color velocity magnitude. We observe that viscoelasticiyty leads to a "jet-like" behavior in the middle of the cavity.  Left: $g \tau = 0.5$. Right: $g \tau = 5*10^{-4}$}
    \label{fig:jet}
\end{figure}

We consider the same geometric setup as before, but now aim to solve the full viscoelastic model (Eq.~\ref{eq:const-viscoel-grow}). We can solve the equations numerically in COMSOL Multiphysics, observing that values of $g \tau$ close to 1 lead to a qualitatively different solution where most of the flow passes through the narrow region near the middle of the cavity (see Fig.~\ref{fig:jet} for an example of the solution).

Based on insights from numerical simulations, we adopt the following ansatz for the stress tensor and velocity field:

\begin{equation}
    \boldsymbol{\tilde{\sigma}} =
    \begin{pmatrix}
         C(x)  & (g y + v_0) B(x)
        \\
        (g y + v_0) B(x) & A(x) + (g y + v_0)^2 D(x)
    \end{pmatrix}
\end{equation}

\begin{equation}
    v = 
    \begin{pmatrix}
       F(x)
       \\
       (g y + v_0) (H(x) + 1 / 2)
    \end{pmatrix}
\end{equation}

The dimensions of the six functions and the boundary conditions they must satisfy are inherited entirely from those already imposed on $\boldsymbol{\tilde{\sigma}}$ and $\mathbf{v}$, and so require no separate specification: since $g y + v_0$ has the dimensions of a velocity, $A$ and $C$ carry the dimensions of stress, $B$ those of stress over velocity and $D$ those of stress over velocity squared, while $F$ has the dimensions of a velocity and $H$ is dimensionless; the conditions closing the system are the no-penetration and partial-slip conditions at the channel walls together with the symmetry of the channel about its midline.

Substituting the ansatz into the Stokes equations, force balance, and incompressibility condition, we obtain a system of equations for the functions $A(x), B(x), C(x), D(x), F(x), H(x)$:

\begin{eqnarray}
    g H + F' = g/2  
    \\
     2 g^2 D' + g B'' = 0 
    \\
     A =\tau (  2 g A H    - F A') + 2 \eta g H
     \\
 g^2 D = \tau(  2 g^2 B H'  - g^2 F  D'  )
     \\
  g B = \tau(   - g^2 B H + g C H'    - g F   B') + \eta g H'
     \\
 C = \tau(   - 2 g C H  - F C') - 2 \eta g H
\end{eqnarray}

Solving these equations perturbatively in the small parameter $g \tau$, we obtain at the first order the effective permeability appearing in Darcy’s law:

\begin{eqnarray}
    k = -\frac{L^2 (3 \eta +L \xi )^3}{3 \eta  \xi  \left(9 \eta ^2+L^2 \xi ^2 (1-2.5 g  \tau )+3 \eta  L \xi  (2- 5.5 g  \tau )\right)}
    \label{eq:darcy-visc}
\end{eqnarray}

that for the sticky boundary $\xi \to \infty$ results in  

\begin{eqnarray}
    k = \frac{k_0}{ 1-2.5 g  \tau }
\end{eqnarray}

Notably, only the combination $g\tau$ enters: a viscoelastic non-growing fluid and a Newtonian growing fluid both have exactly the same permeability as an ordinary Newtonian fluid. Only when growth and viscoelasticity are simultaneously present does the permeability change.

Why does the permeability increase due to the combined effect of growth and viscoelasticity? Consider a viscoelastic, non-growing material being pushed through a channel by a pressure difference. The velocity near the channel walls is zero, so material close to the walls moves through the channel very slowly. When growth is introduced, however, the material near the walls must also move inward---toward the channel center---to satisfy mass conservation. This transfers material from the slow-moving region near the walls into the fast-moving region at the center. The result is an effective ``fast lane'' near the channel center and a ``slow lane'' at the walls; growth allows material that would otherwise remain in the slow lane to enter the fast lane, making the channel effectively less resistive to flow.

For $g \tau > 0.4$ the denominator in Eq.~\ref{eq:darcy-visc} can turn to 0, indicating inapplicability of perturbation theory beyond this regime; however, the equation can still be solved numerically (see Fig.~\ref{fig:jet}). We observe there the strong flow in above mentioned "fast lane" near the center.

\twocolumngrid

\bibliography{references}% Produces the bibliography via BibTeX.

\end{document}